\documentstyle[12pt,aaspp4]{article}

\received{}
\accepted{}
\journalid{}{}
\articleid{}{}
\slugcomment{}
\lefthead{}
\righthead{}

\def\p{\prime}

\begin{document}

\title{The Steepness Ratio Technique: A New Method\\
       to analyze ROSAT All-Sky Survey Extended Sources}

\vskip 30pt
\author{S. De Grandi\altaffilmark{1,2}, S. Molendi\altaffilmark{3}, 
    H. B\"ohringer\altaffilmark{1},\\
    G. Chincarini\altaffilmark{2,4} and W. Voges\altaffilmark{1}}

\altaffiltext{1}{Max-Planck-Institut f{\"u}r extraterrestrische Physik, 
Giessenbachstrasse 1, D-85740 Garching, Germany}
\altaffiltext{2}{Universit\`a di Milano, via Celoria 16, 
    I-20100 Milano, Italy}
\altaffiltext{3}{Istituto di Fisica Cosmica, CNR, via Bassini 15, 
I-20133 Milano, Italy}
\altaffiltext{4}{Osservatorio Astronomico di Brera, via Brera 28, 
I-20121 Milano, Italy}

\begin{abstract}
In this first paper of a series we develop a new technique to analyze
clusters of galaxies observed during the ROSAT All-Sky Survey (RASS).
We call this method the Steepness Ratio Technique (SRT).
The SRT uses the convolution between the real RASS 
point-spread function and the cluster emission profile assumed to be a
$\beta$-model with the $\beta$ parameter fixed to the value of 2/3.
From the convolved source emission profile the SRT extracts total flux and 
extension (i.e., core radius) for each cluster.  
Using the Steepness Ratio (SR) and a control sample of RASS
pointlike sources we are able to assign to each RASS source a 
model-independent probability of extension.
Potential biases arising from the hypotheses of a $\beta$-model emission
profile and from a fixed $\beta$ value are quantified.
Tests with control samples of optically identified
sources extracted from both the ROSAT survey database and from the 
ROSAT-PSPC pointed observations archive
are performed to check the SRT output.
We find that the source properties derived using the SRT on RASS data are
consistent with those determined from ROSAT-PSPC pointed observations.
\end{abstract}


\section{Introduction}
The first part of the ROSAT mission (Tr\"umper 1993) was dedicated
to the  All-Sky Survey 
(RASS; Voges 1992), carried out with the Position Sensitive Proportional 
Counter (PSPC; Briel \& Pfeffermann 1986) as focal plane detector.
Given the vast area explored during the ROSAT survey, in practice the 
whole sky, the RASS data provide a unique opportunity to construct
a large sample of X-ray sources, that can be used for detailed cosmological 
and statistical studies. 
This is exploited in the frame of the ESO Key Program Redshift 
Survey of southern sky clusters of galaxies (B\"ohringer 1994;
Guzzo et al. 1995; De Grandi 1996), which started in 1992 and is
still in progress.
The main aim of this optical follow-up program is the spectral
confirmation and measurement of the recession velocity of
cluster candidates selected from RASS sources in the southern
hemisphere.
This will provide one of the largest samples of X-ray identified
clusters, which will be an ideal tool to map the large scale 
structure in the Universe and investigate the evolution of clusters.

In the soft X-ray band galaxy clusters appear as single entities, as
their emission originates in the thin hot plasma trapped in the
deep cluster potential well. 
The X-ray emission of clusters, predominantly thermal bremsstrahlung,
extends typically over distances on the order of a few Mpc (e.g. Sarazin 1989). 
A study of the properties of these objects in the X-ray band
requires the existence of an algorithm that is able to characterize 
extended sources.
By characterization we mean here the ability to determine 
physical quantities such as the extension and flux of a
source, under conditions where the number of observed photons is low.

Previous techniques applied up to date to RASS data are the Standard 
Analysis Software System (SASS; Voges et al. 1992), developed at the 
MPE institute,
and the Voronoi Tessellation and Percolation (VTP; Ebeling \& Wiedenmann 
1993; Ebeling 1993).

SASS uses two standard sliding window algorithms to detect the sources
observed in the ROSAT survey and applies to each detection a Maximum 
Likelihood (ML) algorithm (Cruddace et al. 1988, Zimmermann et al. 1994) 
in order to derive position, source count rate, existence probability of the 
source, extension and probability of extension.
The ML algorithm is specialized in characterizing pointlike sources as
it takes into account only the point-spread function (PSF) in fitting 
the data.
Further the PSPC point-spread function is assumed by ML, for the purpose
of analytical simplicity, to be a Gaussian which depends on the 
energy and off-axis angle of each photon, therefore the PSF used by ML in the
RASS is a sum of Gaussians (Cruddace et al. 1991).
This is a raw approximation of the real survey PSF (RASS-PSF), because the
sum of Gaussian PSFs does not take into account the important counts
fraction contained in the wings of the real RASS-PSF (of the order of 30\%).
We expect that the ML algorithm, as it is applied in SASS, will lead
to a systematic underestimation of the source counts for both pointlike
and extended sources and also, to an incorrect value for source
extension and associated extension probability 
(see also Cruddace et al. 1991).
In the present work we are mainly interested in the analysis of extended
sources, but we believe that a discussion about the ML effects for
pointlike sources is of great interest (see Appendix A).

VTP is based on concepts completely different to those of SASS.
It analyses the photon distribution directly and does not assume any 
geometrical shape for the source emission profile in the detection process, 
which results in a more accurate determination of the source counts.
However, this initial advantage of VTP is lost in the 
characterization of the detected sources, because the source emission 
profile is modeled using a radially symmetric profile 
(Ebeling et al. 1996).
Moreover, no errors are associated to the VTP extensions (see Ebeling
et al. 1996), this makes interpretation of the results difficult.
Another limitation of VTP as applied in the RASS (i.e. Ebeling 1993;
Ebeling et al. 1996) is that it has only been run in the broad band 
(0.1-2.4 keV). 
However the best ROSAT energy band for the study of 
clusters of galaxies is the hard band (0.5-2.0 keV), because this
is the band in which the signal-to-noise ratio is maximized.

A general criticism of the ML and VTP techniques is the non optimal use
of the survey PSF and the inefficiency in producing a useful extension 
parameter.
Attempts to solve these problems have been made in the past, using
observations of clusters of galaxies  by earlier satellites,
e.g., Lea \& Henry (1988).

The ideal RASS analysis technique for extended sources should:
1) characterize in an adequate  manner the source surface brightness 
profile,
2) use the correct survey PSF,
3) measure the clusters fluxes in the ROSAT hard band
(0.5-2.0 keV),
4) discriminate between pointlike and extended sources in a sensitive
manner, 
5) assign a meaningful physical extension and an 
extension probability to each source, and, 
6) use simple concepts 
and require little processing time for each source.
The present paper is dedicated to the development of a new 
technique that fulfills all these requirements, the Steepness 
Ratio Technique (SRT).

In section 2 we will develop the theory of the SRT starting from the 
convolution between the real RASS-PSF and a source emission profile 
and we will describe how to derive extension (subsection 2.1) and 
flux (subsection 2.2) for RASS clusters of galaxies.
In subsection 2.3 we will check the dependence of the SRT results 
on the parameters of the assumed source emission profile. 

In order to test the reliability of the SRT we will apply, in section 3,
the SRT to control samples of optically identified RASS sources 
and to a sample of ROSAT-PSPC pointed observations of clusters of 
galaxies (subsections 3.2 and 3.3).
In subsection 3.2 we develop a general criterion for discriminating 
between pointlike and extended RASS sources, and a way to associate 
an extension probability to the sources. 
Section 4 will summarize the principal results and conclusions.

In subsequent papers of this series (De Grandi et al. in preparation)
we shall apply the new SRT to the sample of prominent southern sky 
galaxy clusters being studied by the ESO Key Program Redshift Survey
mentioned above, in order to select an X-ray flux 
limited sample of bright clusters and to investigate its properties.
Of prime interest is the X-ray luminosity function.

\section{Theory of the Steepness Ratio Technique (SRT)}
A proper analysis of extended sources requires that the blurring
introduced by the spatial resolution of the observing instruments be 
taken into account correctly.
This is particularly true for the RASS, where the PSF is considerably 
broader than the on-axis PSF for PSPC pointed observations.

During the RASS, the satellite scanned the sky along great circles
perpendicular to the sun position and intersecting the ecliptic poles,
with a progression rate of $\sim 1^o$ per day.
During each orbit, the instrument scanned a region of the sky $\sim 2^o$
wide and $360^o$ long at a constant ecliptic longitude.
As a consequence of this observing strategy each source entered the field
of view of the PSPC once per orbit, and at a slightly different position
during each orbit.
Therefore each source was observed at all off-axis angles in the
detector.
The RASS-PSF has been numerically computed at three different energies 
(0.3, 1.0, 2.0 keV) by averaging the PSF of the (XMA + PSPC)\footnote
{The X-ray mirror assembly (XMA) + PSPC PSF is the point-spread function
obtained by the convolution of the XMA-PSF with the PSPC-PSF.}
PSF over all the detector, weighting the contribution from each off-axis angle 
with the appropriate energy-dependent vignetting factor
(Hasinger, private communication).
This RASS-PSF has been subsequently tested by us on a sample of
bright pointlike RASS sources randomly distributed over the sky.
These tests indicate that the RASS-PSF is in good agreement with the data.
In Figure 1 we compare the on-axis PSF, as available in the EXSAS package
(Zimmermann et al. 1994), with the RASS-PSF both computed at 1 keV. 
We note that the shape of the RASS-PSF cannot be described with a Gaussian
function because of the pronounced wings.
Inspection of Figure 1 shows that a significant fraction of the source counts
for a pointlike source is found at large radii, so that one cannot neglect 
the effect of the RASS-PSF when studying moderately extended sources 
(i.e. $r_c \lesssim 10^\p$).
The ML algorithm implemented in the SASS approximates the survey PSF 
with a sum of Gaussian PSFs and the result is a systematic underestimation of the 
source fluxes both for extended and pointlike objects
(see Appendix A).
In the following we describe a new technique which properly takes into
account the RASS-PSF.

Let us assume a reference system in polar coordinates $(r_{\rm 0},\varphi)$ 
with origin at the point O (see Fig. 2). 
Let $I(r_{\rm 0},\varphi)$ be the emission profile of a source, centered 
at the origin O. 
The flux emitted from an elementary area $dS$ at a projected 
distance $r_{\rm 0}$ from O and at an angle $\varphi$ with respect to the line
$\overline{OP}$, is given by $I(r_{\rm 0},\varphi)r_{\rm 0} dr_{\rm 0} d\varphi$.
Due to the PSF the flux fraction emitted from $dS$ and observed 
at any point P at a distance $r$ from the origin is
\footnote{We use the ROSAT All-Sky Survey PSF (RASS-PSF) computed at the 
energy of 1 keV. As we will show later (section 2.1), the effect of the energy 
dependence of the RASS-PSF on our results is negligible.}
$$I(r_0,\varphi)~ r_0~ dr_0~ d\varphi PSF(d)
\eqno(1)$$
where $d$ is the distance between $dS$ and P.
Therefore, the total flux observed at P is obtained by integrating over the
whole surface the contributions from the elementary areas $dS$:
$$ \tilde I(r)~~ = \int_{0}^{\infty} dr_0 \int_{0}^{2\pi} 
d\varphi ~ I(r_0)~ r_0~ PSF(d),  
\eqno(2)$$
where the distance $d$ is given by
$$d~~=~r_0 \left[ \left({r\over r_0} -\cos\varphi\right)^2
+\sin^2\varphi \right]^{1/2}.
\eqno(3)$$

We assume the surface brightness profile of clusters to be described adequately 
by an isothermal $\beta$-model 
(Cavaliere \& Fusco-Femiano 1976; Jones \& Forman 1984):
$$I(r_0)= I_0 \left[ 1 + \left( r_0\over r_c\right)^2\right]^{-3\beta +1/2},
\eqno(4)$$
which is a function of the central surface brightness, $I_{\rm 0}$, 
the core radius, $r_{\rm c}$ and the $\beta$ parameter.

As most of the RASS sources have a low photon count 
($\lesssim 100$ counts), it is not feasible to determine simultaneously
all the three parameters describing the $\beta$-model.
Previous work (e.g., Jones \& Forman 1984) has shown that $\beta$ is only 
moderately scattered around a mean of $\beta = 2/3$, and therefore we fix 
the $\beta$ parameter at this value.
Substituting eq. (4), for $\beta = 2/3$ into eq. (2) we find
$$ \tilde I(r)~~ = {1\over {2\pi r_c^2}} \int_{0}^{\infty}\int_{0}^{2\pi}
\left[ 1 + \left( r_0\over r_c\right)^2\right]^{-3/2}~
PSF(d)~ r_0 ~dr_0 ~d\varphi 
\eqno(5)$$
where both the $\beta$-model and the PSF are normalized to unity.

A detailed study on the dependence of the technique results upon the 
$\beta$ parameter is given in section 2.3.

In the following we shall illustrate, by means of two examples, 
why in the analysis for an extended source in the ROSAT survey
1) it is crucial not to neglect the effect of the PSF and
2) it is fundamentally important to consider the real RASS-PSF and 
not a Gaussian approximation.

First we solve numerically the integral in eq. (5) using a 
Gaussian PSF with the same FWHM used in SASS analysis of
RASS data ($96^{\p\p}$, corresponding to $\sigma = 40.85^{\p\p}$).
The convolved profile, $\tilde I_g(r)$, is a function of the 
ratio between the core radius 
and the $\sigma$ of the Gaussian PSF: when $r_{\rm c}/\sigma \ll 1$ 
the convolved profile tends to the PSF,
while, when $r_{\rm c}/\sigma \gg 1$, the effect of the convolution becomes
negligible and $\tilde I_g(r)$ tends to the $\beta$-model profile.
In Figure 3 we show the difference between the convolved profile, 
$\tilde I_g(r)$, and the unconvolved $\beta$-model profile, $I(r)$,
normalized to the latter (i.e., $(\tilde I_g(r)-I(r))/I(r)$) for three
values of the $r_{\rm c}/\sigma$ ratio.
In Figure 3a we illustrate the case $r_{\rm c}/\sigma = 0.01$ 
(i.e., $r_{\rm c}/\sigma \ll 1$), here the convolved profile approximates
the PSF which has a very different shape from the $\beta$-model profile.
Consequently the quantity $(\tilde I_g(r)-I(r))/I(r)$ represented in
Figure 3a is very different from zero.
On the contrary in Figure 3c, where $r_{\rm c}/\sigma = 100$
(i.e., $r_{\rm c}/\sigma \gg  1$), the convolved profile approximates
the $\beta$-model and the quantity $(\tilde I_g(r)-I(r))/I(r)$ is always
very close to zero.

In the intermediate cases, when $r_{\rm c}$ and $\sigma$ are comparable, the
convolved profile, $\tilde I_g(r)$, differs significantly from both the
unconvolved and the PSF profiles.
Since the effect of the convolution is to broaden the profile and both the
convolved and unconvolved profiles are normalized to unity, we have
that the convolved profile is weaker than the unconvolved one at small radii
and vice versa is stronger for large radii (see Fig. 3b).
Therefore, in studying RASS sources with core radii of the same order of 
magnitude of $\sigma$ the convolution of the PSF must be taken into account.

In the second example we compare the Gaussian PSF convolved profile 
with the RASS-PSF convolved profile.
Figure 4a shows the product of $\tilde I(r)~ r$ as a function of the 
radius, for a ratio $r_{\rm c}/\sigma = 3$.
The solid line is the product computed for the RASS-PSF convolved profile,
whereas the dashed line is the product computed using the Gaussian PSF.
As the convolved profiles are normalized to unity, i.e.
$ 2\pi \int_{0}^{\infty} r~ \tilde I(r)~ dr ~\equiv~ 1$,
the areas below the curves in Figure 4a are equal.
The spreading effect of the convolution on the source emission profile is 
larger in the case of the RASS-PSF (solid line in Fig. 4a) than in the case 
of a Gaussian PSF (dotted line).
Figure 4b shows the ratio $(\tilde I(r)-\tilde I_{\rm g}(r))/\tilde I_{\rm g}(r)$ 
as a function of the radius $r$, where $\tilde I(r)$ is the profile
convolved with the RASS-PSF and $\tilde I_{\rm g}(r)$ the profile convolved
with a Gaussian PSF.
The profiles differ significantly and the ratio is larger than
10\% for radii larger than 4$^\p$, indicating that it is not 
possible to approximate the real PSF with a simple Gaussian function.

\subsection{Extension}
Once the RASS-PSF convolved $\beta$-model profile has been derived it can be used
to extract information on the extension of RASS sources.
Since the objects observed during the survey are characterized by
low statistics, we choose to compare the convolved profile (5) with the data
using integrals of $\tilde I(r)$.
In particular, we integrate eq. (5) in a circle with a radius of $3^{\p}$
and in an annulus bounded by the two radii $3^{\p}$ and $5^{\p}$,
obtaining the counts fractions
	$$ C(3^\p)~=~ 2\pi \int_{0}^{3^\p} r~ \tilde I(r)~ dr;$$
	$$C(5^\p-3^\p)~=~ 2\pi \int_{3^\p}^{5^\p} r~ \tilde I(r)~ dr.
	\eqno(6)$$
Now we consider the ratio between the two above integrals:
$$ SR~ \equiv~ {{C(5^\p-3^\p)}\over C(3^\p)}.
\eqno(7)$$
We name this quantity the steepness ratio ($SR$), as it is a
measure of the slope of the convolved profile $\tilde I(r)$.
There is a strong analogy between the steepness ratio, which we will
use in deriving the extension of the sources, and the hardness ratio
used as a measure of the spectral slope.
In both cases photon counts are accumulated into two bins, as a result 
of the limited statistics of the sources.

As $SR$ is a monotonic function of the core radius $r_{\rm c}$, (Fig. 5)
it may be used to obtain the core radius from the observed steepness ratio, 
$SR_{\rm obs}$.
The $SR_{\rm obs}$ is measured from the survey data as the ratio between the 
source counts falling into a $3^{\p}$ radius circle and the source counts
falling into an annulus bounded by the two radii of $3^{\p}$ and $5^{\p}$:
$$ SR_{obs} =~ {{cts(5^\p-3^\p)}\over cts(3^\p)}.
\eqno(8)$$
The uncertainty in measuring $SR_{\rm obs}$ is computed applying the usual 
formulae for error propagation, using the errors in the two independent 
quantities $cts(3^\p)$ and $cts(5^\p-3^\p)$.
The $SR - r_{\rm c}$ curve is applied to compute the errors in the derived 
core radius, using the known errors in $SR_{\rm obs}$.

In Figure 5 we also illustrate the case of the $SR - r_{\rm c}$ curve computed
from an unconvolved $\beta$-model profile (dotted curve). 
The ratio $(SR_{\rm \tilde I} - SR_{\rm I})/SR_{\rm I}$ between the RASS-PSF convolved 
and unconvolved $SR - r_{\rm c}$ curves is shown in Figure 6.
As can be clearly seen in Figure 6 the two curves, drawn in Figure 5, are quite 
different and only for core radii larger than $11^\p$ do they differ by less 
than 1\%, while at $2^\p$ the difference is $\sim 50$\%.

To examine the energy dependence of the $SR - r_{\rm c}$ curve, we have computed
different $SR - r_{\rm c}$ curves using the RASS-PSF at 0.3 keV and 2.0 keV.
The ratios between the curves, computed at different energy, is never
greater than 2\% for $r_{\rm c} \lesssim 10^{\p\p}$ and it is always less 
than a few 0.1 percent at larger core radii.
We conclude that the energy dependence of the curve $SR - r_{\rm c}$ is negligible.

At core radii $r_{\rm c} \lesssim 10^{\p\p}$ the convolved profile (eq. 5) approaches
the RASS-PSF.
Consequently, as shown in Figure 5, the $SR$ becomes insensitive to the core 
radius and approaches the value 0.15, which is characteristic of pointlike 
sources (see eq. 7).
For very large $r_{\rm c}$, the dominant profile is that of the $\beta$-model
and the photons distribution between 0$^\p$ and 5$^\p$ is practically flat
so that the $SR - r_{\rm c}$ curve tends to the constant value 1.78
(i.e., the ratio of the areas of the annulus $(5^\p-3^\p)$ and the circle
of radius $3^\p$).
As a corollary at large ($r_{\rm c} > 1000^{\p\p}$) and small 
($r_{\rm c} < 40^{\p\p}$) radii, small variations of $SR$ lead to large 
variations in the core radius, thus in these regions the $SR - r_{\rm c}$ 
curve is not suitable to measure the core radius.
On the other hand when the derivative of $SR$ with respect to $r_{\rm c}$ 
is large (i.e. in the range $60^{\p\p}\lesssim r_{\rm c}\lesssim 900 ^{\p\p}$),
the curve provides a good estimation of the core radius.
For a typical cluster, with physical core radius of 250 kpc (e.g., Bahcall 
1975), this corresponds to a redshift range of $0.009 < z < 0.2$ 
($H_{\rm 0} = 50$ km/s/Mpc).

A different choice for the radii of the circle and the annulus,
results in a new $SR - r_{\rm c}$ curve, and changes the range in which
$r_{\rm c}$ can be derived adequately.
Inspection of Figure 5 shows that $SR = 1$ for $r_{\rm c}\sim 200^{\p\p}$. 
Had we chosen a significantly larger circle and annulus,
$SR$ would have been equal to 1 for a larger core radius and therefore
would be inadequate to characterize barely extended sources.
A significantly smaller circle and annulus would allow us to characterize such
sources, but would provide poor information for more extended objects.

\subsection{Total Source Counts}
In the case of a pointlike source it is possible to compute the source counts
by counting the events within a circle of a fixed extraction radius and
then correcting for the counts falling outside that radius 
using the PSF.
In the case of an extended sources this correction must take into account
the source extension, using the following procedure.

Consider a source for which we have obtained the core radius using the SRT.
As the RASS-PSF convolved emission profile in eq. (5) is normalized so that
$2\pi \int_{0}^{\infty} r~ \tilde I(r)~ dr \equiv 1$,
the total source counts are given by 
$$ cts_{tot} = {cts~(5^{\p}~)~ F},
\eqno(9) $$
where $cts~(5^{\p})$ are the observed source counts in 5$^{\p}$, and 
$$ F = {1\over 2\pi \int_{0}^{5^\p} r~ \tilde I(r)~ dr}.
\eqno(10) $$
\smallskip\noindent
The correction factor $F$ is a function dependent only on $r_{\rm c}$, 
$\beta$(=2/3) and the PSF.
As the steepness ratio (eq. 7) is likewise a function of these three 
quantities, it follows that $F$ is a function of $SR$, as shown in Figure 7.

The $SR - F$ curve can be used to compute the errors in the derived correction 
factor $F$, from the errors on $SR_{\rm obs}$.
The uncertainties ascribed to the total source counts, $cts_{\rm tot}$, 
are computed applying the usual formulae for error propagation, using the
errors in the two independent quantities $cts~(5^{\p}~)$ and $F$.
In Figure 7 we also show the $SR - F$ curve computed using an unconvolved
$\beta$-model profile (dotted curve). 
The ratio $(F_{\rm \tilde I} - F_{\rm I})/F_{\rm I}$ between the RASS-PSF 
convolved and unconvolved curves, drawn in Figure 8, shows significant 
differences in the derived total flux.
Only for core radii larger than about
14$^\p$ do they differ by less than 5\%, while at 3$^\p$ the difference is
$\sim 12$\%.

To study how the energy dependence of the RASS-PSF impacts the 
$SR - F$ curve 
we have recomputed the $SR - F$ curve using the 0.3 and 2.0 keV RASS-PSF.
We find that the ratio between the nominal curve, computed at 1 keV,
and either of the other two curves is never greater than 3\% and
therefore conclude that the energy dependence of the RASS-PSF has a
negligible effect upon the computation of the total source counts.

At small core radii, $r_{\rm c} < 10^{\p\p}$, the correction factor F tends 
to 1.049 (Fig. 7), i.e. the inverse of the RASS-PSF integral between 0 and
$5^\p$.
This limit is the correction that has to be applied for pointlike sources,
this implies that SRT can also be used to compute total source counts
for pointlike sources.
On the other hand at very large core radii, $r_{\rm c} > 1000^{\p\p}$,
the correction factor F increases indefinitely, so that, as in the case
of the core radius, the SRT is not applicable to objects 
with very large core radii ($r_{\rm c} \gtrsim 1000^{\p\p}$).

\subsection{Dependence of the SRT Results upon the ${\rm \beta}$ Parameter}
One of the fundamental hypotheses which the SRT is based on is the
assumption of a fixed value for the $\beta$ parameter, $\beta = 2/3$.
We investigate now how this hypothesis could affect the measure 
of the core radius and of the total counts of a RASS source.
In previous work (e.g., Jones \& Forman 1984) it has been found that 
the value of the $\beta$ parameter for galaxy clusters ranges between 
$\sim 3/5 - 4/5$.
Therefore, we recompute the $SR - r_{\rm c}$ and $SR - F$ curves 
for the two limiting values  3/5 and 4/5 for $\beta$, and compare them with the
curves computed for $\beta = 2/3$.

\subsubsection{Core Radii}
The $SR - r_{\rm c}$ curves computed for a RASS-PSF convolved $\beta$-model
for $\beta$ equal to 3/5, 2/3 and 4/5 are plotted in Figure 9. 
The $SR$ of the three curves tend to the same limits for small and large 
$r_{\rm c}$, and the core radius derived for $\beta = 2/3$ always lies between
the values of $r_{\rm c}$ obtained for $\beta = 3/5$ and $4/5$.
Using the curves in Figure 9 we compute the following ratios:
$$ \Delta r_c(\beta = 3/5,\beta = 2/3)\equiv
{{r_c(SR,\beta=3/5) -r_c(SR,\beta=2/3)}\over r_c(SR,\beta=2/3)},$$
$$\Delta r_c(\beta = 4/5,\beta = 2/3)\equiv
{{r_c(SR,\beta=4/5) -r_c(SR,\beta=2/3)}\over r_c(SR,\beta=2/3)}.
\eqno(11)$$
In Figure 10 these ratios are plotted as a function of the core radius
obtained for $\beta = 2/3$. 
As we will show in the next section (3.), values of $r_{\rm c}(SR,\beta=2/3)$ 
below 50$^{\p\p}$ are not interesting, because for these values is not 
possible to distinguish between pointlike and extended RASS sources.
Figure 10 shows that the largest difference in the core radius ($\sim 50$\%)
is found for $r_{\rm c}$ = 50$^{\p\p}$, and the differences decrease as the core
radius increases, falling to $\sim 7$\% at $r_{\rm c} \sim 1000^{\p\p}$.
The difference in the core radii found at small values, near $50^{\p\p}$,
depends upon the radii chosen for the circle and the annulus used in eq. 7.

\subsubsection{Total Source Counts}
We consider now how different $\beta$ parameter values could affect the total 
source counts.
Analogous to the procedure shown in the previous section we compute the 
$SR - F$ curves for $\beta$ = 3/5, 2/3 and 4/5, yielding the results shown
in Figure 11.
We define the ratios:
$$ \Delta F (\beta = 3/5,\beta = 2/3)\equiv
{{F(SR,\beta=3/5) - F(SR,\beta=2/3)}\over F(SR,\beta=2/3)},$$ 
$$\Delta F (\beta = 4/5,\beta = 2/3)\equiv
{{F(SR,\beta=4/5) - F(SR,\beta=2/3)}\over F(SR,\beta=2/3)}. 
\eqno(12)$$
These ratios are plotted in Figure 12 as a function of the core radius
$r_{\rm c}(SR,\beta=2/3)$.
At small core radii these ratios tend to zero, as the correction factor $F$ 
approaches that found for pointlike sources.
The ratios increase with increasing core radius, reaching a maximum value of 
$\sim$ 30\% when the core radius is about $1000^{\p\p}$.

We conclude that the assumption $\beta = 2/3$ could lead to an error in the
derived total source counts ranging from 0 to 30\%.

\section{Applying the SRT to Control Samples}
The reliability of the SRT has been tested using different control samples
of optically identified sources observed in the RASS.
These samples represent an alternative to the use of simulated RASS fields. 
Indeed a ROSAT observation, carried out during the survey, is the
result of complex scan operations and often is influenced by random events,
for example the PSPC switching off during the passage through intense charge 
particles zones (i.e. the South Atlantic Anomaly), attitude problems occurring 
during some scans, and scan reversals every 30 days, which were made 
to avoid earth occultations.
For these reasons realistic simulations of RASS observations are
hard to plan. 
Therefore we preferred to use control samples of real RASS sources, which 
take into account directly all the observational difficulties present in 
the survey.

Our control samples are: 
1) the EMSS sample (Gioia et al. 1990, Maccacaro et al. 1994) reobserved 
in the RASS, 
2) an all-sky sample of 262 bright RASS stars selected by correlating
the SASS source list (Voges 1992) with the SIMBAD database at the Centre 
de Donnes Astronomiques de Strasburg, and yielding more than 100
SASS counts in the hard band (0.5-2.0 keV), and   
3) a sample of 26 Abell clusters (Abell et al. 1989) extracted from the
public archive of deep ROSAT-PSPC pointed observations.

\subsection{Data Analysis}
We analyze RASS fields of $2^o\times 2^o$, centered on the SASS sources 
positions.
The data are produced by merging all the RASS scans at a position in the sky.
An automatic analysis procedure has been developed that uses the spatial 
analysis techniques available within the EXSAS (Zimmermann et al. 1994) 
package.
The algorithm: 
1) applies a local source detection (LD) algorithm to the binned data,
\footnote{If the detection algorithm finds a secondary emission peak
within a $10^\p$ radius from the primary peak, then the source is
flagged.}
2) produces a ``swiss-cheesed'' image which is obtained 
by removing the events from $5^\p$ 
circles centered on the detected source positions,
\footnote{in the case of the central source, which may be extended,
we cut out a circle with a radius of $24^\p$, to avoid contamination
of the background from the halo of a possibly extended source.}
3) computes a background map of the field by performing a 2-dimensional 
cubic spline fit of the ``swiss-cheesed'' image,
4) applies the Maximum Likelihood algorithm (ML) to the unbinned data, 
deriving a position, an existence likelihood, a count rate, an extension 
and an extension likelihood for the source at the center of the field,
5) extracts the source counts in a $3^\p$ circle and in an ($5^\p-3^\p$) 
annulus centered on the position found by the ML algorithm and
6) applies the SRT deriving an extension ($r_{\rm c}$), an extension probability
(see section 3.2) and the total source counts ($cts_{\rm tot}$).
This procedure is applied independently using each of the three ROSAT energy 
bands, i.e. the broad (0.1-2.4 keV), soft (0.1-0.4 keV) and hard (0.5-2.0 keV)
bands.

\subsection{Separating Extended from Pointlike Sources} 
We consider the EMSS subsample of 30 RASS bright sources, obtained by
selecting 
objects with more than 100 counts in the ROSAT hard band (0.5 - 2.0 keV)
within a $5^\p$ radius circle around the ML position.
In Figure 13 we show the distribution of core radius for this subsample.
The pointlike sources (dashed histogram) have core radii smaller that 
$\sim 50^{\p\p}$, with the only exception indicated by the black bar.
This object is the BL-Lac MS1207.9+3945 which is located $\sim 5^\p$
from one of the brightest AGN in the sky NGC4151. The spatial resolution
of the RASS does not allow to separate MS1207.9+3945 from NGC4151.
The extended sources, clusters of galaxies and galaxies (white histogram),
clearly show a distribution that is different from that of pointlike
sources.
The SRT core radii distribution for the control sample of 262 bright stars 
identified using SIMBAD is shown in Figure 14.
The distribution decreases rapidly at radii smaller than 20$^{\p\p}$ 
and goes to zero at about 60$^{\p\p}$.

In the following we develop a general criterion for discriminating 
between pointlike and extended RASS sources.
Such a criterion is extremely useful in the preidentification phase
of RASS sources, i.e. once a RASS source is recognized as being extended,
it can belong only to a few astrophysical classes of objects.
Moreover, in the case of sources localized beyond the galactic
plane, the most probable classes of extended objects are galaxies and 
clusters of galaxies.

We consider the control sample of bright stars 
and the EMSS pointlike objects identified as stars or AGN.
In order to have a fair sample of very bright pointlike sources,
we selected in both samples only those sources with more than 100 hard 
band (0.5-2.0 keV) counts within a $5^\p$ radius circle.
In total we obtain 286 bright pointlike sources.

For these sources we define the observed distribution of $SR$ as
$dn_{\rm occ} (SR)/dSR$, where $dn_{\rm occ}$ is the occurrence number for
$SR$ in the interval $dSR$.
Normalizing to the total number of objects, $n_{\rm occ}$, we obtain
the observed probability density:
$$ {dP_{obs}\over dSR} ~=~ {1\over n_{occ}} ~ {dn_{occ}(SR)\over dSR}.
\eqno(13)$$
A cubic spline-fit algorithm applied to the distribution (13) 
leads to a continuous representation of the probability density function,
$dP/dSR$, and the distributions $dP_{\rm obs}/dSR$ and $dP/dSR$ are shown 
in Figure 15.
The two distributions peak at a low value of $SR$ around 0.17, and then
fall rapidly to zero for values of $SR$ greater than 0.3.

We use now the $dP/dSR$ curve to associate an extension probability for each
RASS source.
The case of sources with negligible errors is considered first, after which
errors are included.
Consider a source with an observed steepness ratio, $SR_{\rm obs}$.
The probability for a pointlike source to have a steepness ratio greater
or equal to $SR_{\rm obs}$, is given by:
$$ P(\geq SR_{obs})~=~  \int_{SR_{obs}}^\infty {dP\over dSR}~ dSR.
\eqno(14)$$
As shown in Figure 15, the greater the value of $SR_{\rm obs}$ 
the smaller is the probability $P(\geq SR_{\rm obs})$ for a source with
$SR \geq SR_{\rm obs}$ to be pointlike.
Equation  (14) is valid if the error, $\sigma_{SR_{\rm obs}}$, is 
small with respect to $SR_{\rm obs}$.
However if $\sigma_{\rm SR_{\rm obs}}$ is not negligible, we must take into 
account the probability distribution, $dG/dSR$, associated with the measured 
$SR_{\rm obs}$.
A Gaussian distribution of errors is assumed:
$$ {dG\over dSR} ~=~ {1\over \sigma_{SR_{obs}} \sqrt{2\pi}}
~ \exp \left[ - {1\over 2} \left( {SR - SR_{obs}\over \sigma_{SR_{obs}}}
\right)^2 \right] 
\eqno(15)$$
The analog of eq. (14) is obtained weighting the probability
$P(\geq SR_{\rm obs})$ over the probability distribution $dG/dSR$ of the
source, giving
$$P(\geq SR_{obs}) ~=~ \int_0^\infty dSR^\p~ {dG\over dSR^\p}~
\int_{SR^\p}^\infty dSR~ {dP\over dSR}. 
\eqno(16)$$
When $\sigma_{\rm SR_{\rm obs}}/SR_{\rm obs}\rightarrow 0$ the 
Gaussian given in eq. (15) 
tends to a Dirac function, and eq. (16) 
turns into eq. (14). 

We can now use the probability defined in eq. (16) 
to test our null order 
hypothesis (i.e. that an X-ray source belongs to a population of pointlike objects).
Therefore, we call ``extended'' those sources having an associated 
probability, $P(\geq SR_{\rm obs})$, as defined in eq. (16), 
smaller than 0.01.
In the case of a source with an error $\sigma_{\rm SR_{\rm obs}}$ negligible
with respect to $SR$, a probability of 0.01 corresponds to $SR = 0.334$.

\subsection{Checking the SRT Source Count Rates}

\subsubsection{Checking Count Rates for Pointlike Sources}
To check the accuracy of the SRT in determining the source count 
rates, we consider all the 131 EMSS pointlike objects identified as 
stars or AGN detected in the RASS (i.e., we do not apply any cut in
source counts).
Moreover, we use the source counts instead of the count rates, because 
counts show more clearly statistical effects.

By integrating the RASS-PSF at 1 keV between 0$^\p$ and 5$^\p$ we find that
the expected counts fraction falling in a circle of 5$^\p$ radius from the  
RASS-PSF peak is 95.3\%.
Therefore the ratio $(cts_{\rm tot} - cts(5^\p))/cts(5^\p)$, 
where $cts_{\rm tot}$ and $cts(5^\p)$ are respectively the total source 
counts (eq. 9) and the counts in $5^\p$, expected for a pointlike source
is 0.05.
Averaging the above ratio over the whole distribution of pointlike EMSS
sources, we find that it is consistent with the expected value
($0.06\pm 0.015$).

As we pointed out previously SRT is a technique specifically developed 
to characterize RASS sources when the photon statistics is low.
In the following we investigate how SRT performs as a function of the 
signal-to-noise ratio of detection.

In the analysis of the EMSS pointlike sources we find that for a fraction
of them (23/131) SRT is not able to compute core radius nor total source 
counts.
This occurs when the statistical fluctuations of the counts in circles of 
$3^\p$ and $5^\p$ radii give unphysical $SR$ (eq. 7), e.g., when the 
counts within $3^\p$ are larger than those within $5^\p$.
To quantify this effect we define the signal-to-noise ratio as the ratio
between the source counts measured within a circle of $5^\p$ radius and the
square root of the total counts (i.e., source + background) measured in the
same circle, and we study the capability of SRT to characterize RASS sources 
as a function of the signal-to-noise.
We find that SRT does not miss any EMSS pointlike sources, due to an
unphysical $SR$ value, when the signal-to-noise ratio is larger than 4.
This corresponds to $\sim 20$ source counts within $5^\p$.
For signal-to-noise ratios equal to 3, 2 and 1 the technique fails 
respectively in the $8\%\pm6\%$, $45\%\pm14\%$ and $33\pm19\%$ of cases.
The source photon counts into $5^\p$ associated with signal-to-noise of 
3, 2 and 1 are small, i.e. $\sim 15$, 10 and 5 respectively.

Finally we investigate how well SRT estimates the flux of pointlike 
sources as a function of the signal-to-noise ratio.
We consider again the ratios $(cts_{\rm tot} - cts(5^\p))/cts(5^\p)$
defined above, and we compute the weighted means of these ratios 
averaged on bins of signal-to-noise.
We find that the means are always consistent with the aspected value,
within less than a few 0.1\%.
Therefore the count rates estimated by SRT are not affected by systematical
errors at any signal-to-noise ratio.

\subsubsection{Checking Count Rates for Extended Sources}
In this section the SRT count rates obtained from RASS data are compared
with count rates measured from ROSAT-PSPC pointed observations of a sample 
of 26 Abell clusters of galaxies, which we have analyzed using the EXSAS
package.
We sum the counts in a circle of appropriate radius centered on the cluster 
emission peak and subtract the background estimated in a region void of sources.
As the majority of the clusters are significantly extended, we
consider vignetting by weighting each photon with the PSPC response function 
at the appropriate off-axis angle and energy.
We obtain the count rates by dividing the source counts by the exposure time 
of the observation.

In Figure 16 we show the relation between the RASS-SRT and the pointed
observations count rates.
The uncertainties in the count rates are 1-sigma errors, and both count
rates are computed in the hard band (0.5-2.0 keV).
The correlation between the two independent quantities is good and is valid for 
a large dynamic range, i.e. count rates range from $\sim 0.3$ to 4 cts/s.
The SRT leads to significantly different count rates in the case of
three clusters only (A2877, A3376 and A3266). 
A detailed analysis of the pointed data showed that, in all three 
cases, the real emission profile of the source differed significantly 
from a $\beta$-model profile with $\beta = 2/3$, because of the very 
peculiar morphology or asymmetry of the cluster.

Note that, the hypothesis of radial symmetry of the sources, implemented in the 
SRT, does not introduce any systematic bias in the measured count rates.
Very extended RASS clusters which are not fitted well using the $\beta$-model 
with $\beta = 2/3$, can exhibit widely differing steepness ratios
values (see eq. 8), because the photons in 3$^\p$ and 5$^\p$ can be 
distributed in many ways, depending on the cluster morphology.
Under these circumstances the SRT may lead to either an under- or 
overestimation of the count rate.
In the case of clusters with extension comparable with the FWHM 
of the RASS-PSF, the inhomogeneities are blurred significantly by the 
RASS-PSF, so that deviations of the intrinsic source profiles from the 
radial symmetry are less important.
As pointed out in De Grandi (1996) and we will show in the next paper
of this series (De Grandi et al. in preparation) the majority of clusters 
observed in the ROSAT survey have extensions comparable with the size of 
the RASS-PSF. 
Therefore SRT will derive an incorrect count rate for a minority
of significantly extended clusters only.

Figure 16 shows that no correction to the SRT count rate is necessary, 
unlike the techniques used to date. 
For example the VTP technique (Ebeling \& Wiedenmann 1993), which computed
the RASS count rates in the ROSAT broad band only, 
is affected significantly by the presence of X-ray pointlike sources 
within clusters and therefore requires correction for contamination
of the RASS count rates on a statistical basis (Ebeling et al. 1996).

Finally we quantify how the SRT count rate measure holds for RASS 
extended sources as a function of the signal-to-noise ratio of detection.
We use the signal-to-noise ratio as defined in section 3.3.1 and 
the EMSS sources identified with galaxies and galaxy clusters detected 
in the hard band RASS data (21 objects).
We find that in no case SRT fails to characterize the sources, due to an 
unphysical value of $SR$, for signal-to-noise $\gtrsim 3$ (corresponding 
to about $10-20$ source counts within $5^\p$).
Moreover, we have checked the validity of the SRT count rates as a 
function of the signal-to-noise for the extended sources using the 
independent data set of the ROSAT-PSPC pointed observations for 
the 26 Abell clusters mentioned above.
For each Abell cluster we compute the ratio
$(cts_{\rm tot} - cts_{point})/cts_{point}$,
where $cts_{\rm tot}$ and $cts_{point}$ are respectively the RASS total source
counts (eq. 9) and the counts measured from the pointed observations, and
than we compute the averaged mean of these ratios in signal-to-noise bins.
We verify that for signal-to-noise $\gtrsim 5$ (corresponding to about
$25-35$ source counts within $5^\p$) the mean value of the ratio is always
within $\sim$ 1\% of the expected value.

\section{Conclusions}
The new steepness ratio technique developed in this paper is
particularly suitable to characterize the extended RASS sources
in conditions of low signal-to-noise ratio.
For the first time in the RASS clusters of galaxies analysis, 
this technique convolves the source surface brightness 
profile and the real point-spread function of the ROSAT survey.
We have verified that a Gaussian approximation of the PSF,
as implemented in the standard RASS analysis software (SASS) 
leads to an underestimation 
of the source counts (i.e. fluxes) both for pointlike and extended sources.
In the SRT we assume a model describing the surface brightness profile 
of clusters of galaxies, namely the $\beta$-model with the $\beta$ parameter 
fixed to the value of 2/3. 
In previous work (e.g., Jones \& Forman 1984) it has been found that the 
$\beta$ parameter for galaxy clusters is moderately scattered around a mean 
of 2/3. Therefore we have performed a detailed study of the dependence 
of the SRT results upon the $\beta$ parameter, which shows that a fixed 
$\beta = 2/3$ could in extreme cases lead to errors on the total source 
counts of no more than 30\%.

We tested the reliability of the SRT applying the new technique on 
control samples of optically identified X-ray sources.
The most important test has been the comparison of the SRT count rates 
computed from RASS data with the count rates computed from deep
ROSAT-PSPC pointed observations, both measured in the hard band 
(0.5-2.0 keV), for a sample of Abell clusters of galaxies.
We found good agreement, implying that no corrections to the SRT count
rates are necessary.

All the tests we have performed lead to the conclusion that the
steepness ratio technique is a robust estimator of the flux for the 
RASS clusters of galaxies and pointlike sources.
Using the steepness ratio (SR) and a control sample of RASS
pointlike sources we were able to assign to each
RASS source a model-independent probability of extension.

In the following papers of this series we will apply the SRT to a defined
sample of galaxy clusters candidates, obtained from the ESO Key Program 
Redshift Survey of southern sky clusters, 
in order to select an X-ray flux-limited sample 
of bright clusters and to investigate in detail the X-ray properties of 
this sample.

\acknowledgments

SD would like to thank R. Cruddace for a critical reading of the
manuscript.
SD acknowledges also useful discussions with G. Zamorani, C. Izzo and
A. Edge.
The authors would like to thank G. Hasinger for having provided
the numerical RASS-PSF.
This work has been performed within the framework
of the ESO Key Program Redshift Survey of southern ROSAT clusters, and the
contribution from the project team is gratefully acknowledged.

\appendix

\section{Appendix}
In this section we discuss the results of the Maximum Likelihood (ML) 
method (Cruddace et al. 1988) applied to the survey data through the 
Standard Analysis Software System (Voges et al. 1992) at the 
MPE.

We select from the EMSS sample (Gioia et al. 1990, Maccacaro et al. 1994) 
a subsample of 131 pointlike sources (i.e. optically 
identified AGN, BL-Lac and stars) which were reobserved in the RASS and
derive for them the ML source counts 
from the merged data as described in section 3.
As the counts fraction falling into a circle of 5$^\p$ radius
from the the RASS-PSF peak is 95.3\%, we choose to compare the ML source 
counts of pointlike objects, $cts_{\rm ML}$, to the source counts inside 
5$^\p$, $cts(5^\p)$ (see Fig. 17). 
For both methods the source counts are measured in the broad band (0.1-2.4 keV).
From Figure 17 we note two important effects:
1) the ML method systematically underestimates the source counts, and
2) the source counts underestimation is a function of the
counts number and it is larger for the weaker sources.
The ratio, $(cts_{\rm ML} - cts(5^\p))/cts(5^\p)$, averaged over the
whole distribution is $0.22\pm 0.01$.

The ML algorithm leads to an underestimation of the source counts
of the pointlike objects because it uses a sum of Gaussian PSFs
as the RASS-PSF.
It is possible to explain the effect observed in Figure 17 by means of
simple considerations.
The surface brightness profile of a pointlike RASS source is described 
by the RASS-PSF, that has pronounced wings (see Fig. 1).
If the source is weak, only the central part of the RASS-PSF is emerging 
from the local background level.
Therefore, the ML method, that uses a sum of Gaussian PSFs to describe
the source, is forced to fit with greater precision the part of the 
RASS-PSF with the better statistics, namely the core.
Hence, the PSF used by the ML is not able to fit the wings
of the real RASS-PSF and consequently underestimates the counts
of a weak source.

On the other hand the surface brightness profile of a bright pointlike 
source emerges more distinctly from the local background, and the statistics 
in the wings is much better than in the case of a weak source.
The ML algorithm includes an attempt to analyze extended sources, by which a 
test is made to see whether a Gaussian surface brightness profile improves
the fit.
This has the unfortunate result that the algorithm, working with a sum of
Gaussian PSFs, interprets the wings of the real PSF as an extension of 
the source.
Consequently, while this procedure makes a more accurate estimate of the
flux from a strong pointlike source, at the same time it assigns a false
extension. 
This is confirmed in Figure 18, which for the sample of bright stars
described in section 3 shows a correlation between the total counts and the 
extension likelihood derived by ML.

\clearpage

\figcaption
{The ROSAT All-Sky Survey PSPC point-spread function (solid line) is 
compared with the on-axis PSPC point-spread function (dotted line). 
Both PSFs are computed at the energy of 1.0 keV, and normalized to unity.}

\figcaption
{Sketch of the geometry involved in computing the convolution between
a source emission profile $I$ centered in the origin O, and the point-spread
function. }

\figcaption
{Ratios $(\tilde I_g(r) - I(r))/I(r)$ as a function of the radius $r$.
The quantities $\tilde I_g(r)$ and $I(r)$ are respectively, the PSF 
convolved and unconvolved $\beta$-model emission profiles.
The PSF used in the convolution is a Gaussian function with
$\sigma = 40.85^{\p\p}$.
In (a) the convolved and unconvolved profiles are computed for a 
ratio $r_{\rm c}/\sigma \ll 1$, in (b) for $r_{\rm c}/\sigma \approx 1$ and 
in (c) for $r_{\rm c}/\sigma \gg 1$. }

\figcaption
{(a) Product $\tilde I(r)~ r$ as a function of the radius $r$.
The quantity $\tilde I(r)$ is the PSF convolved $\beta$-model profile
computed for a ratio $r_{\rm c}/\sigma = 3$.
The solid line is the profile convolved with the RASS-PSF,
whereas the dashed line is the profile convolved with a Gaussian
PSF with $\sigma = 40.85^{\p\p}$.
The two profiles are normalized to unity, i.e. the areas below the
solid and dotted lines are equal.
(b) Ratio $(\tilde I(r) - \tilde I_{\rm g}(r))/\tilde I_{\rm g}(r)$ 
as a function of the radius $r$.
$\tilde I(r)$ and $\tilde I_{\rm g}(r)$ are obtained from convolutions 
with the real RASS-PSF and a Gaussian PSF respectively. }

\figcaption
{The steepness ratio, $SR$ as a function of the core radius, $r_{\rm c}$.
The solid line is the curve computed using the $\beta$-model convolved
with the RASS-PSF, and the dotted line is computed using a simple 
$\beta$-model profile.
Vertical and horizontal lines show how to derive the core radius (solid 
lines) from the steepness ratio and its errors (dashed lines). }

\figcaption
{The difference between the $SR$ value computed by the RASS-PSF 
convolved $\beta$-model, $SR_{\rm \tilde I}$, and that obtained from 
the unconvolved $\beta$-model, $SR_{\rm I}$ (cfr. Fig. 5), plotted 
as a function of the core radius $r_{\rm c}$. }

\figcaption
{The dependence of the correction factor F upon the steepness ratio $SR$.
The solid line is the curve computed using the RASS-PSF convolved 
$\beta$-model profile, and the dotted line that obtained using a simple 
$\beta$-model profile.
Vertical and horizontal lines show how to derive the correction
factor (solid lines) and its errors (dashed line) from steepness ratio.}

\figcaption
{The differences between the correction factor F, computed by the RASS-PSF 
convolved $\beta$-model $F_{\rm \tilde I}$, and that obtained from the 
unconvolved $\beta$-model, $F_{\rm I}$ (cfr. Fig. 7), plotted
as a function of the core radius $r_{\rm c}$.}

\figcaption
{The $SR - r_{\rm c}$ curve computed for a RASS-PSF convolved $\beta$-model
emission profile for different values of $\beta$.
The solid line is for $\beta = 2/3$, the dotted one for $\beta = 3/5$ and 
the dashed one for $\beta = 4/5$.}

\figcaption
{The ratios $\Delta r_{\rm c}$, defined in eq. (11), as a function of
the core radius $r_{\rm c}$.
The solid line shows $\Delta r_{\rm c}(2/3,2/3)$, the dotted line 
$\Delta r_{\rm c}(3/5,2/3)$ and the dashed line $\Delta r_{\rm c}(4/5,2/3)$.}

\figcaption
{The dependence of the correction factor F upon the steepness ratio $SR$. 
over a range of parameter $\beta$: $\beta$ = 3/5 (dotted line), 2/3 
(solid line) and 4/5 (dashed line).}

\figcaption
{The ratio $\Delta F$, defined in eq. (12), as a function of the 
core radius $r_{\rm c}$.
The solid line shows $\Delta F(2/3,2/3)$, the dotted line $\Delta F(3/5,2/3)$ 
and the dashed line $\Delta F(4/5,2/3)$.}

\figcaption
{SRT core radii distribution for the 30 brightest EMSS sources.
The dashed histogram shows the core radii distribution of pointlike sources, 
whereas the unshaded histogram shows the distribution of clusters of 
galaxies and galaxies.
The source colored in black is the BL-Lac MS1207.9+3945 which is located
$\sim 5^\p$ from the very bright AGN NGC4151.}

\figcaption
{The distribution of core radii measured with the SRT for the RASS sample
of bright stars.}

\figcaption
{The distribution of the observed probability density, $dP_{\rm obs}/dSR$, 
for the pointlike sources with the steepness ratio, $SR$. 
The solid line shows the $dP/dSR$ curve obtained applying a cubic spline-fit 
to the observed distribution $dP_{\rm obs}/dSR$.}

\figcaption
{A comparison between the count rates of 26 bright Abell clusters measured 
from ROSAT-PSPC pointed observations and the corresponding SRT count rates 
obtained from the RASS data.
Both count rates are computed in the hard band (0.5-2.0 keV).}

\figcaption
{Comparison between the source counts measured within a 5$^\p$ radius 
circle from the emission peak and the ML source counts.
The sample comprises EMSS pointlike objects observed in the RASS. 
Both were measured using the RASS data in the broad band (0.1-2.4 keV).}

\figcaption
{Correlation of the extension likelihood for a sample of bright RASS stars
with the source count, using the results of the ML analysis.
Both quantities were measured using the RASS data in the hard band 
(0.5-2.0 keV).}

\clearpage

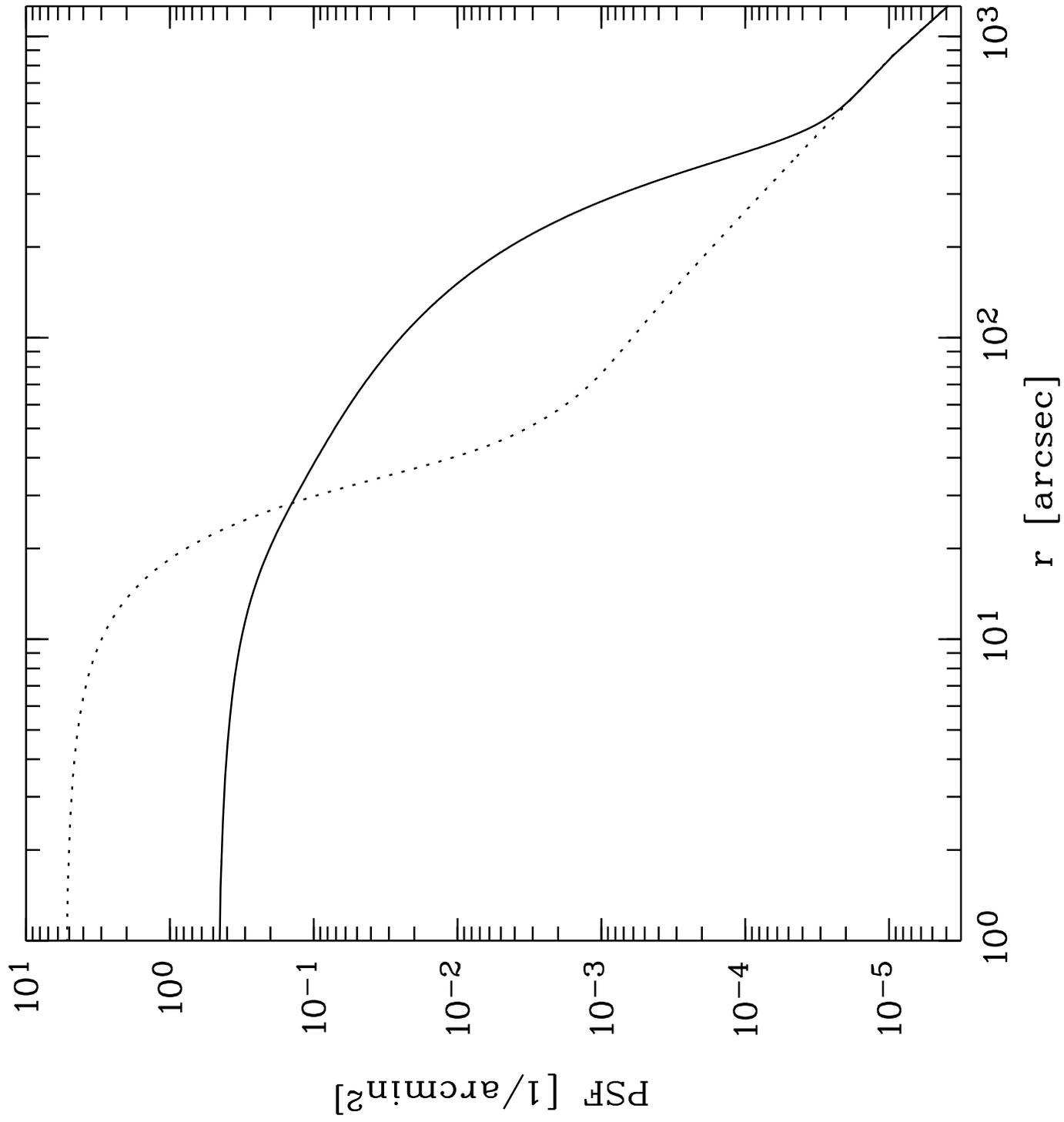
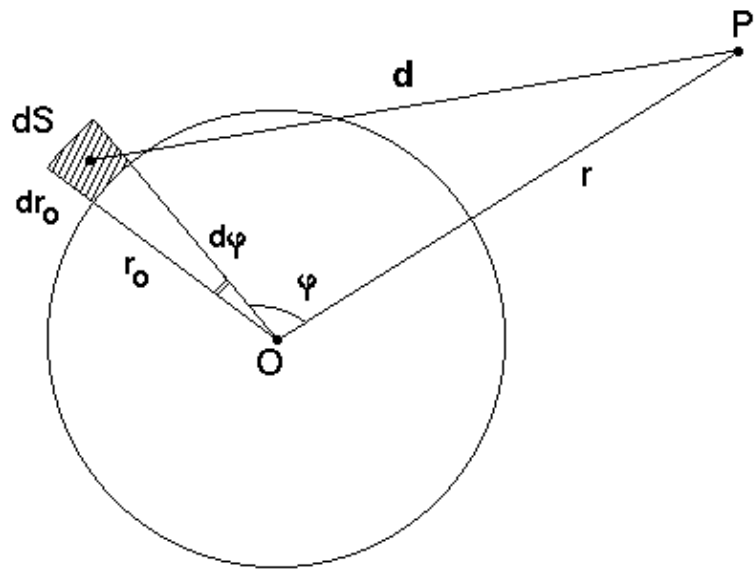
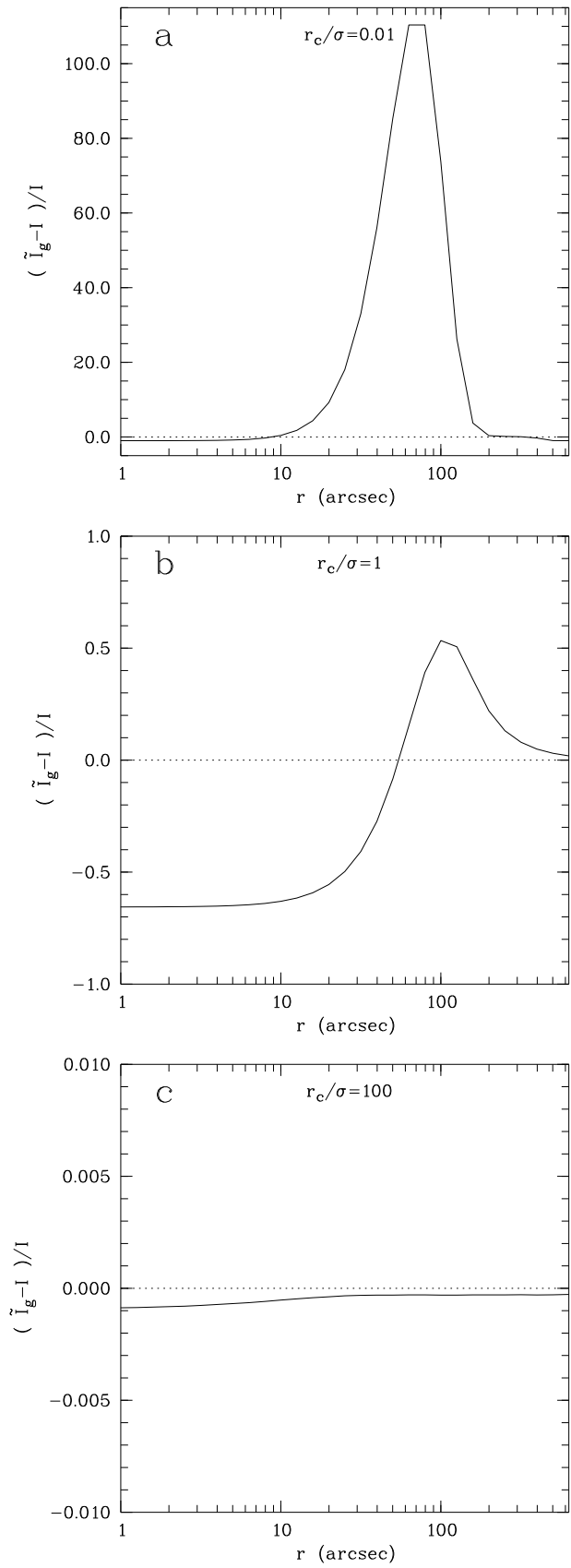
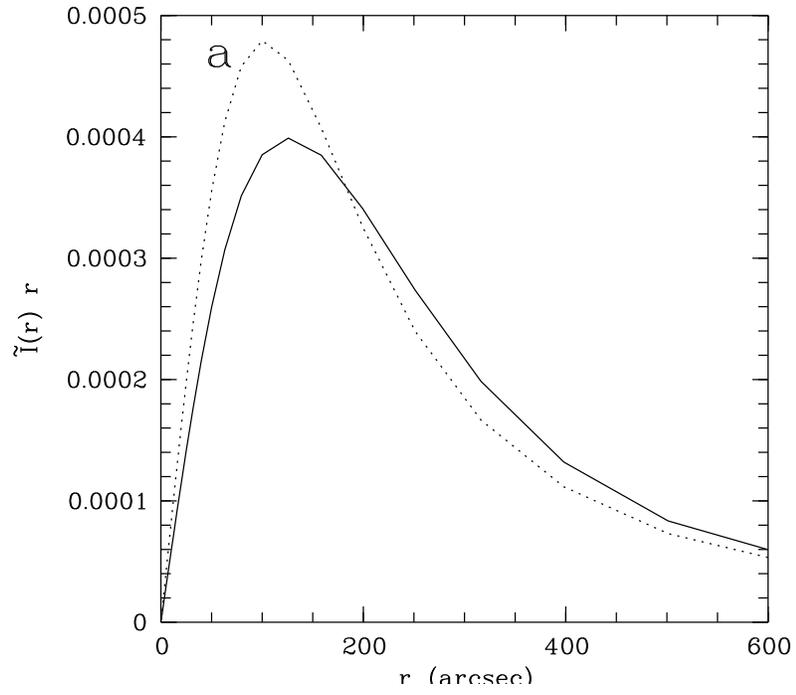
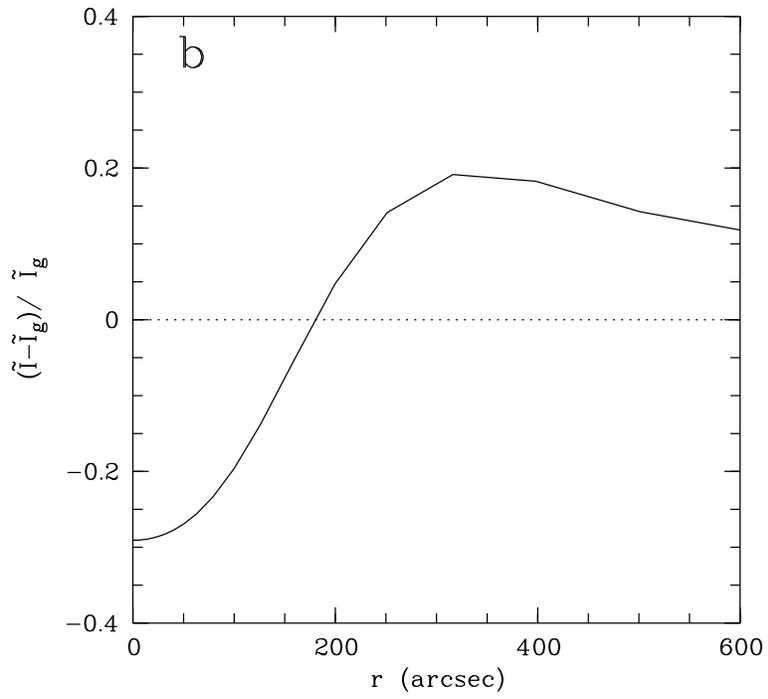
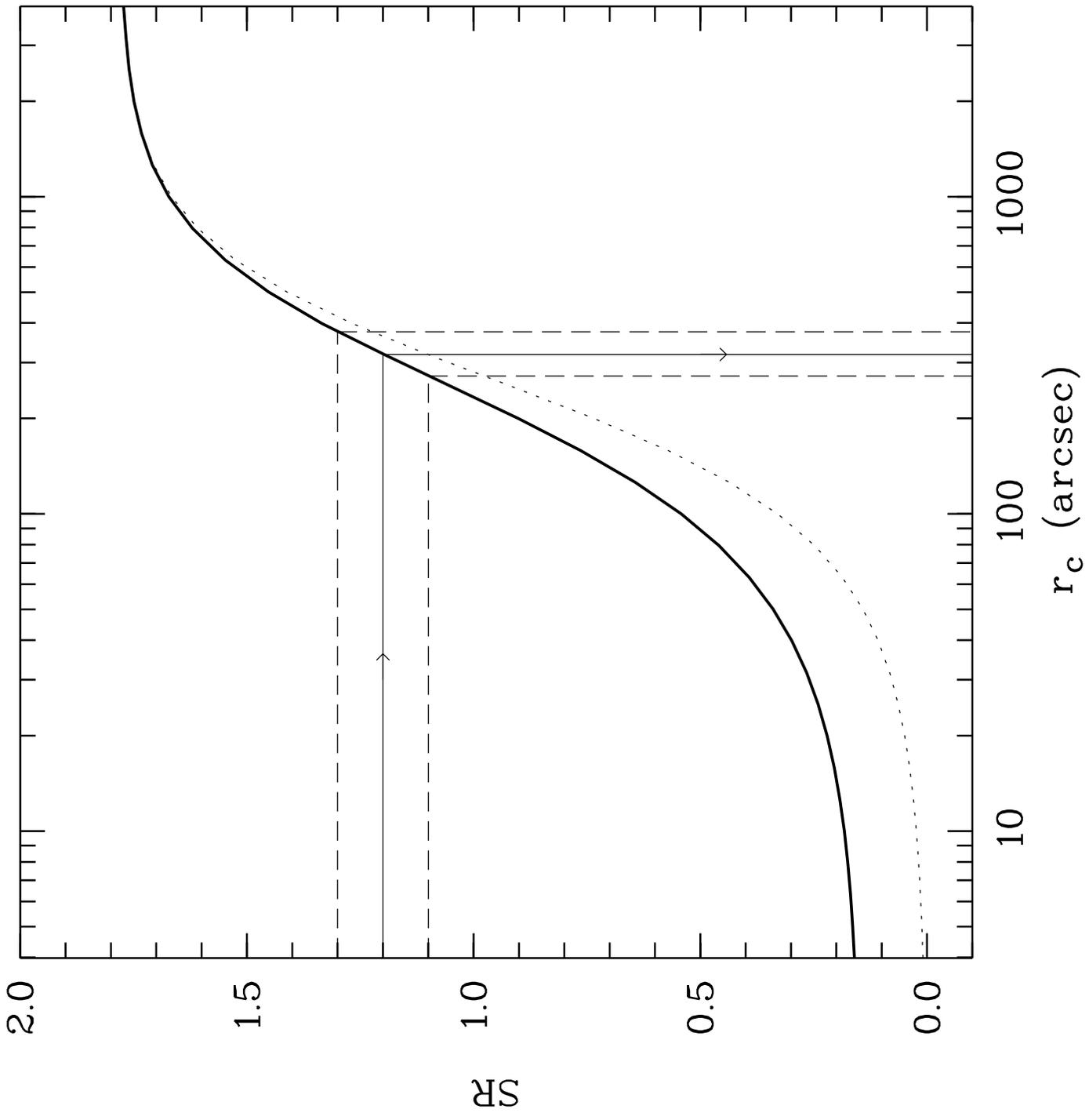
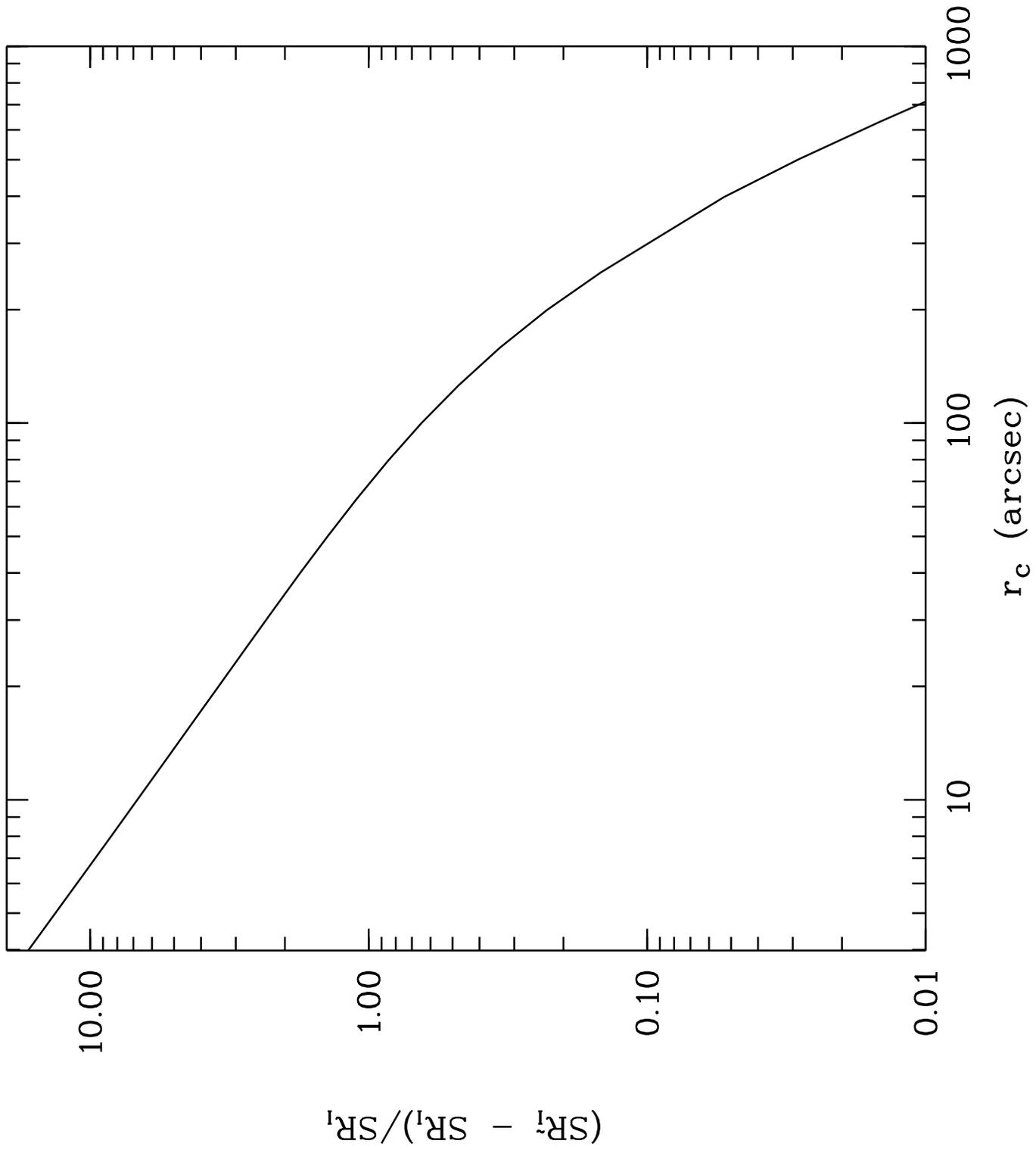
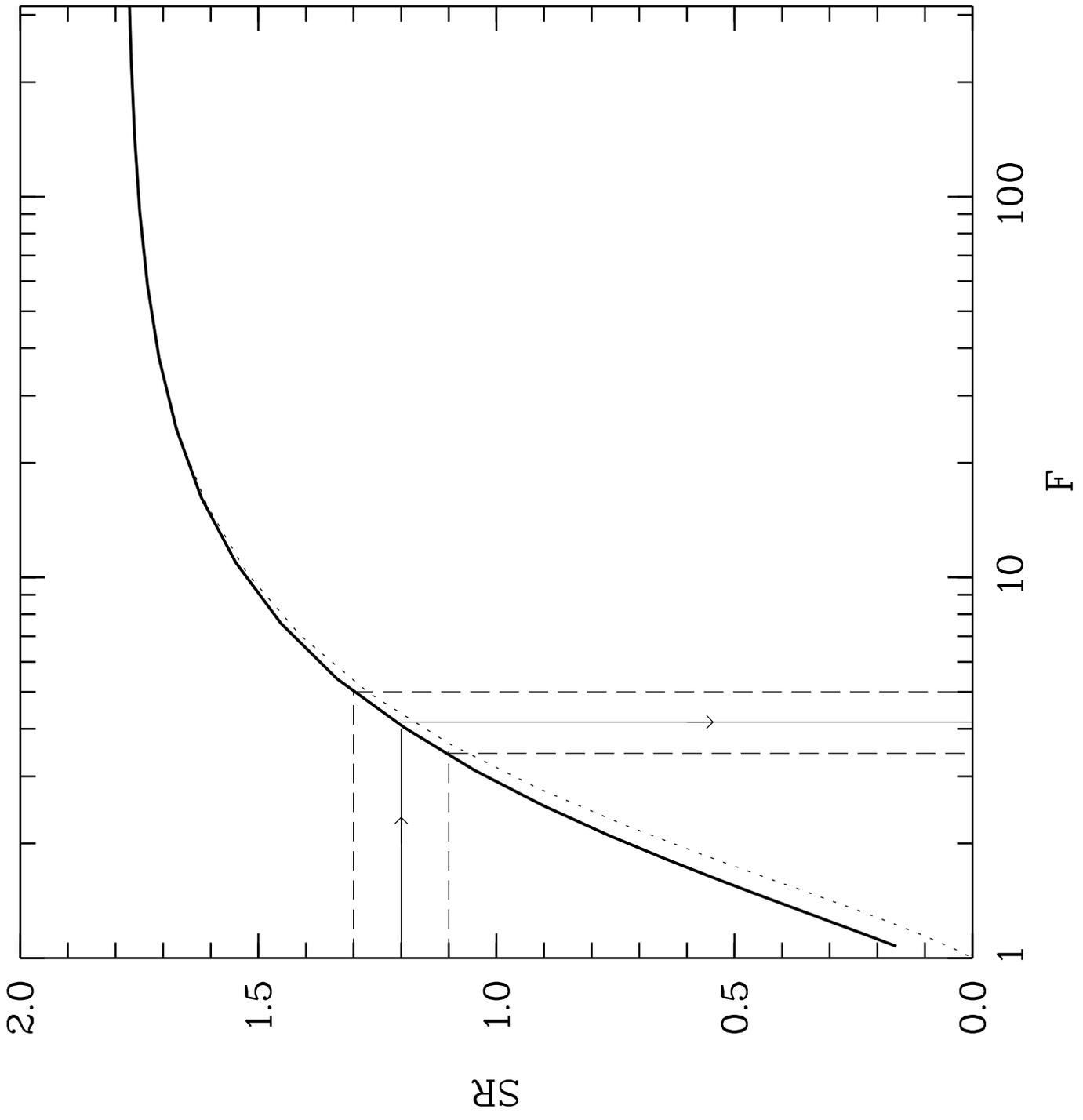
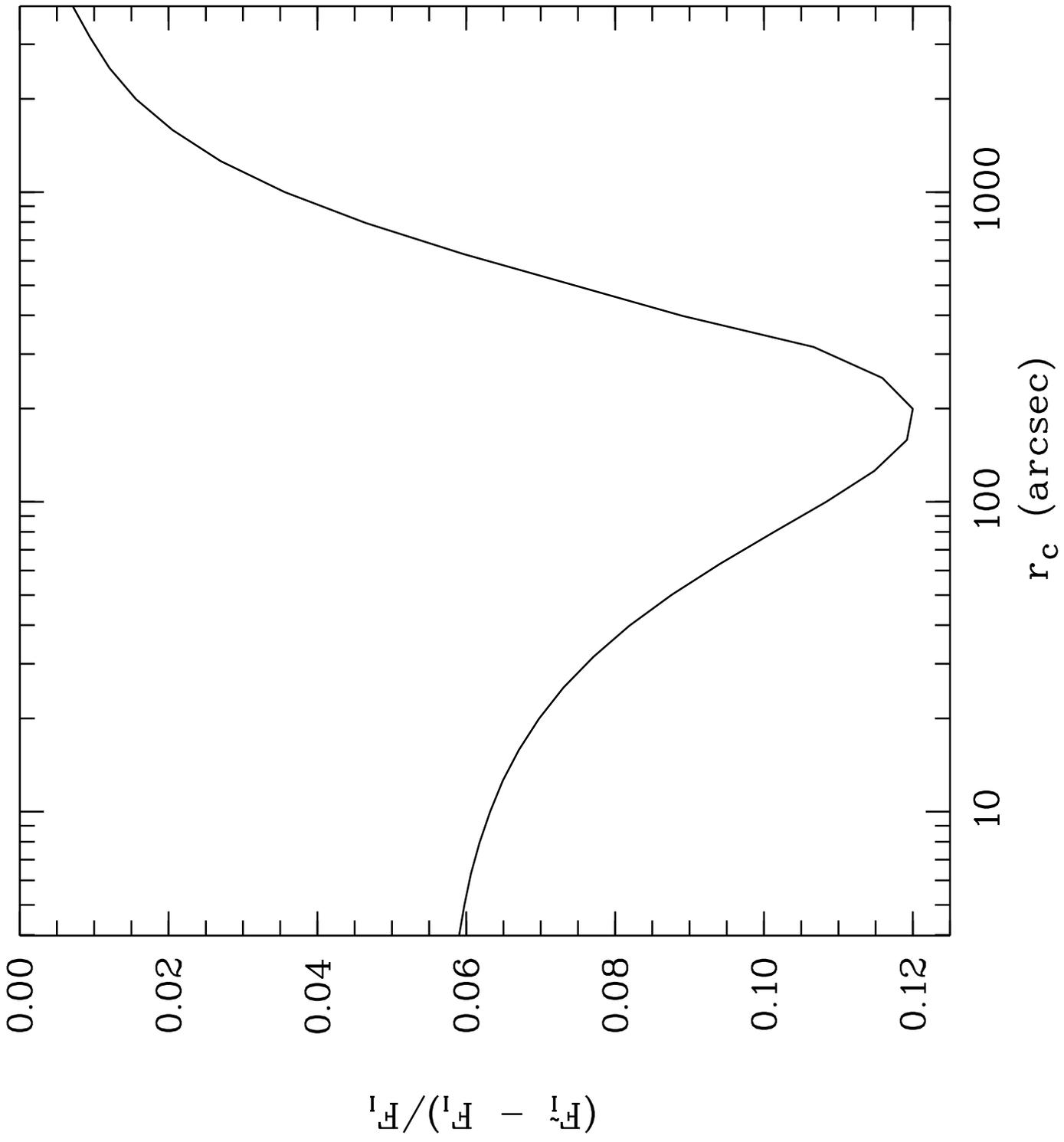
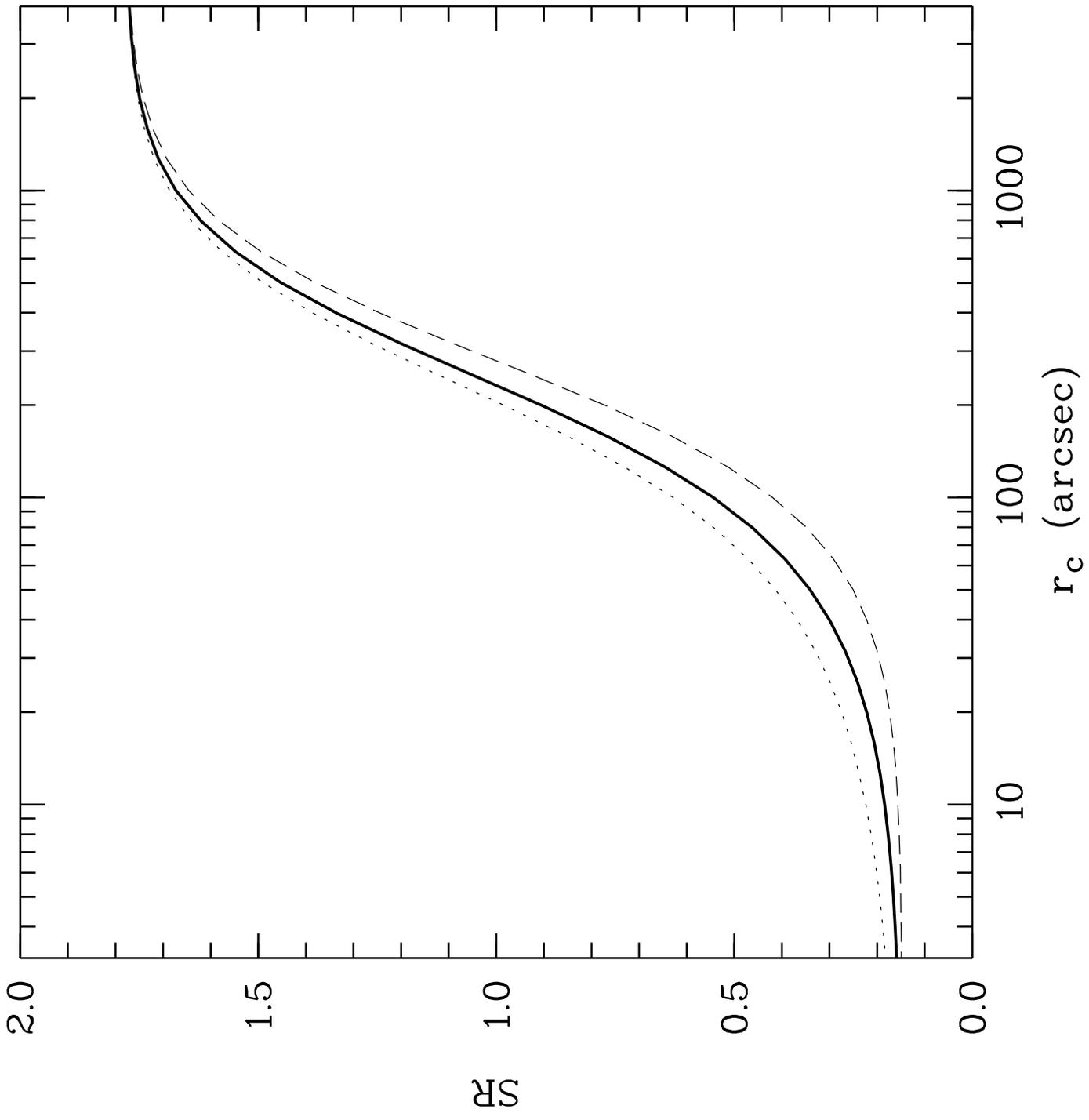
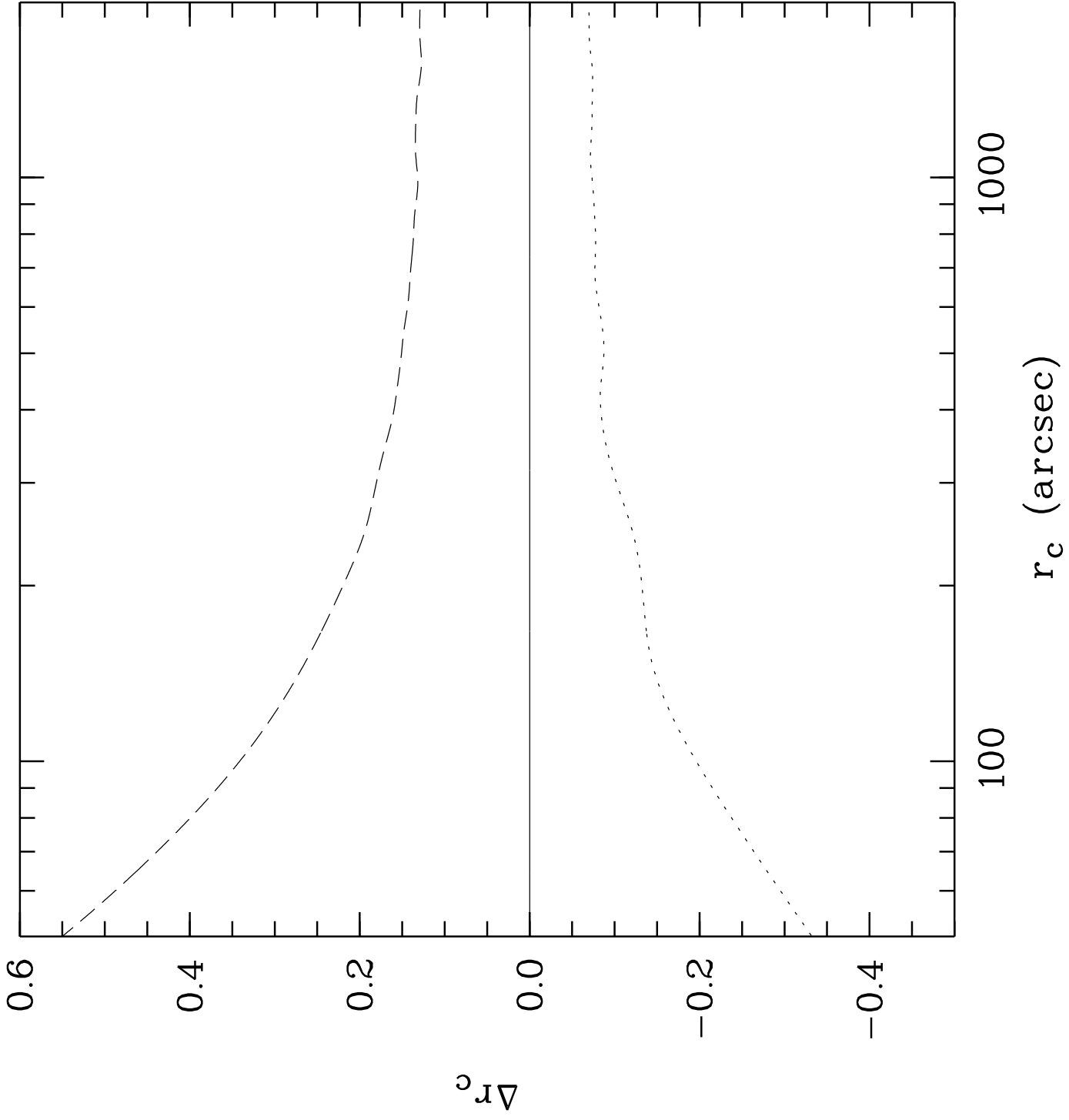
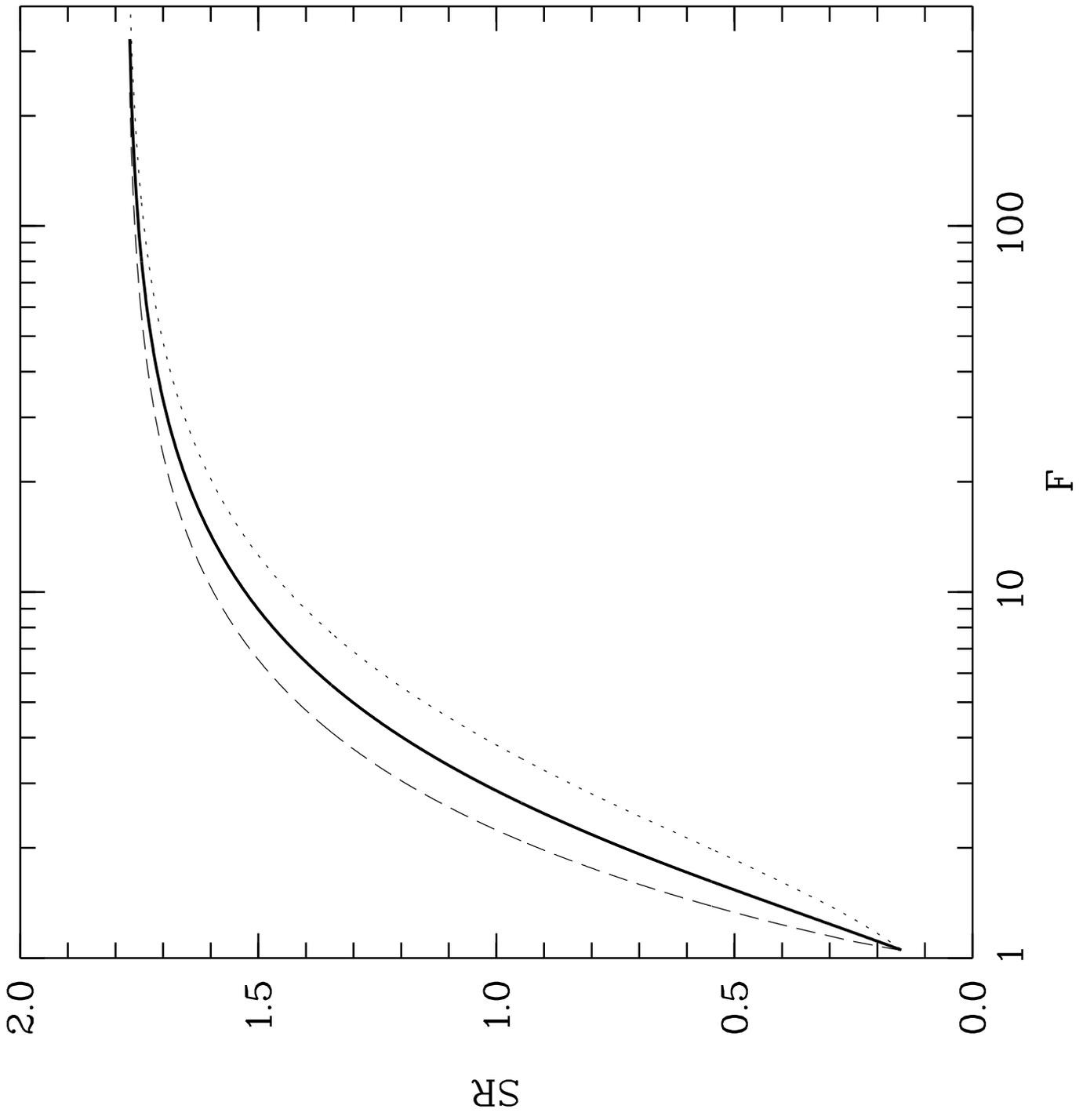
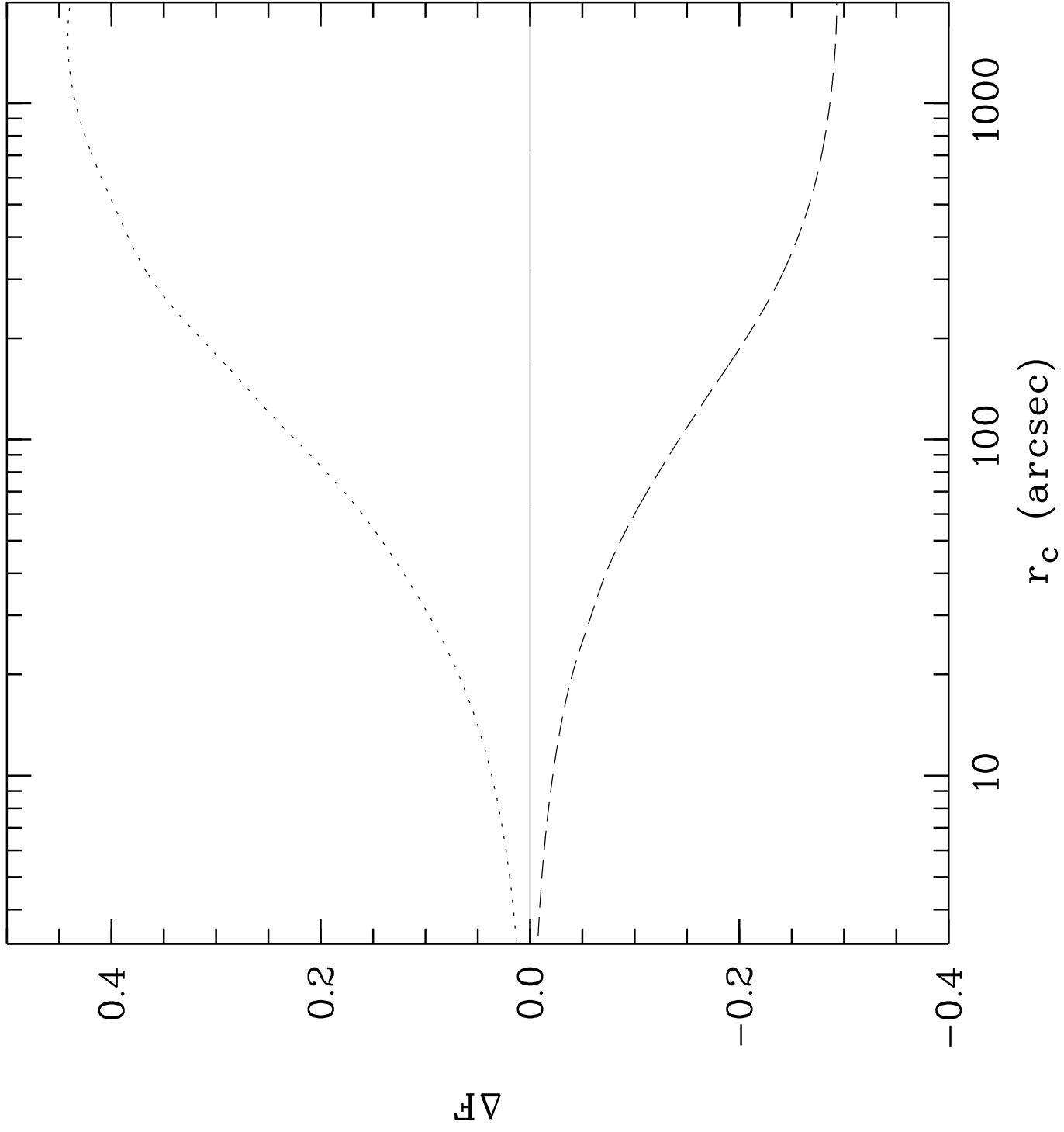
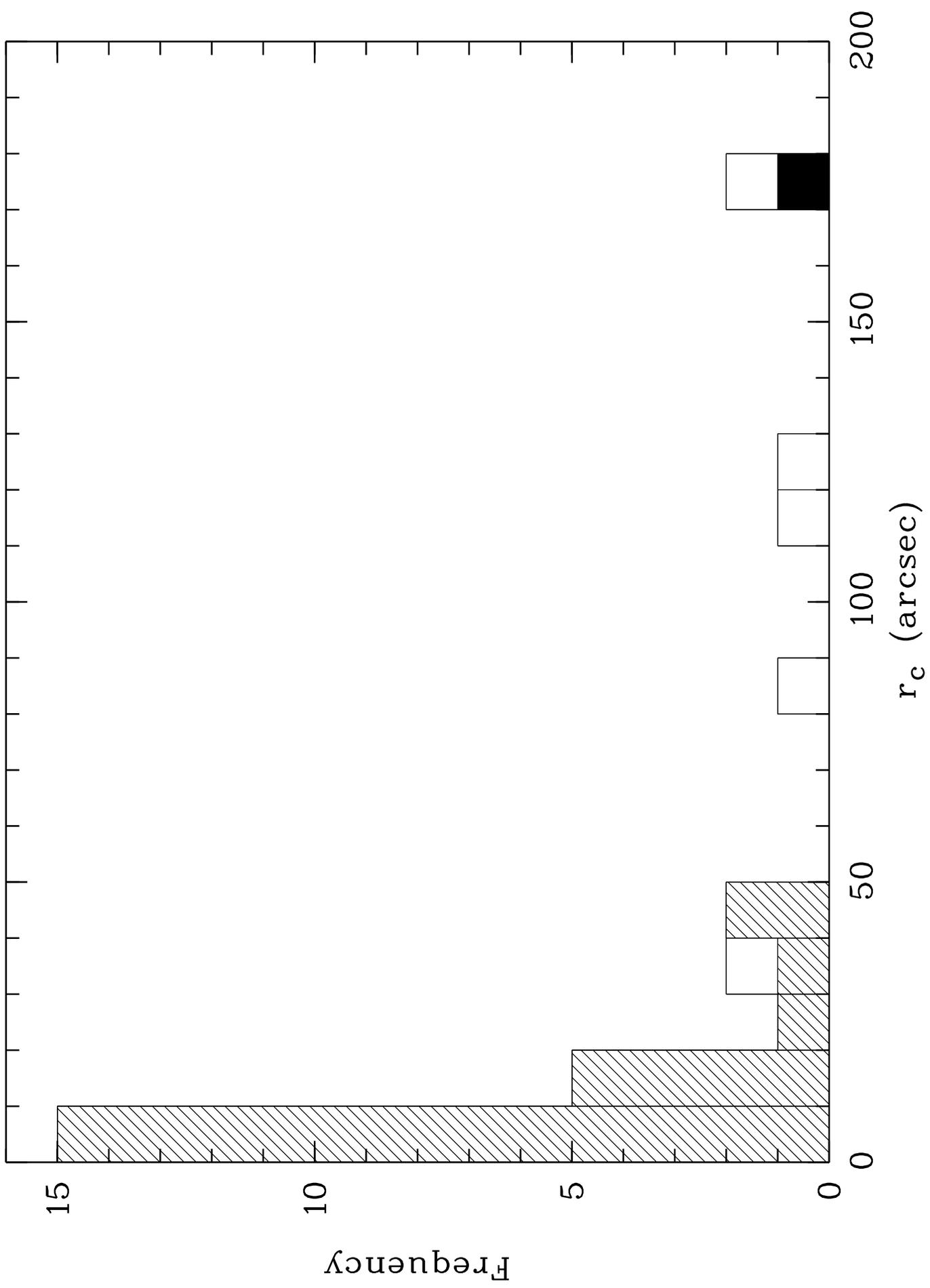
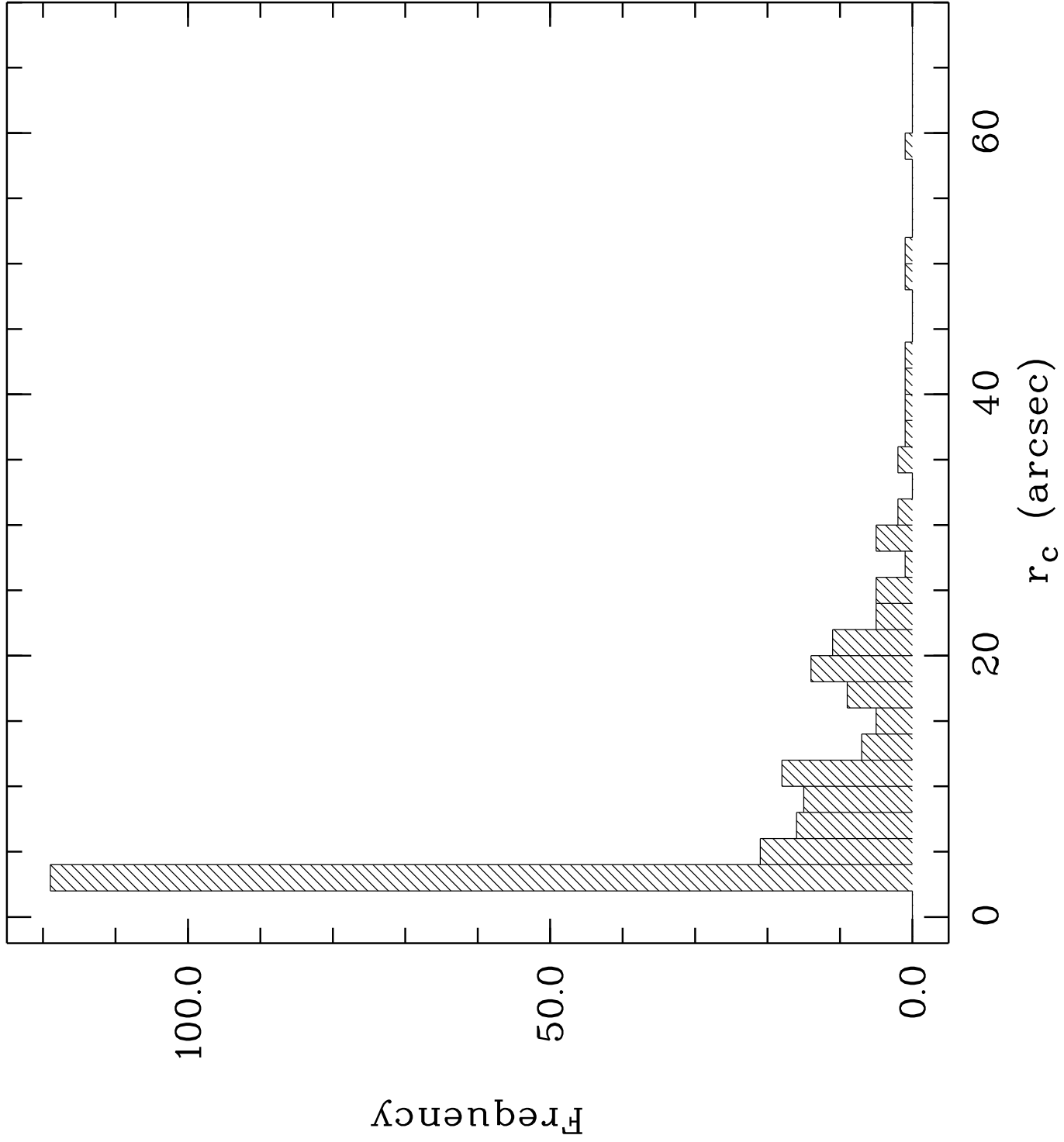
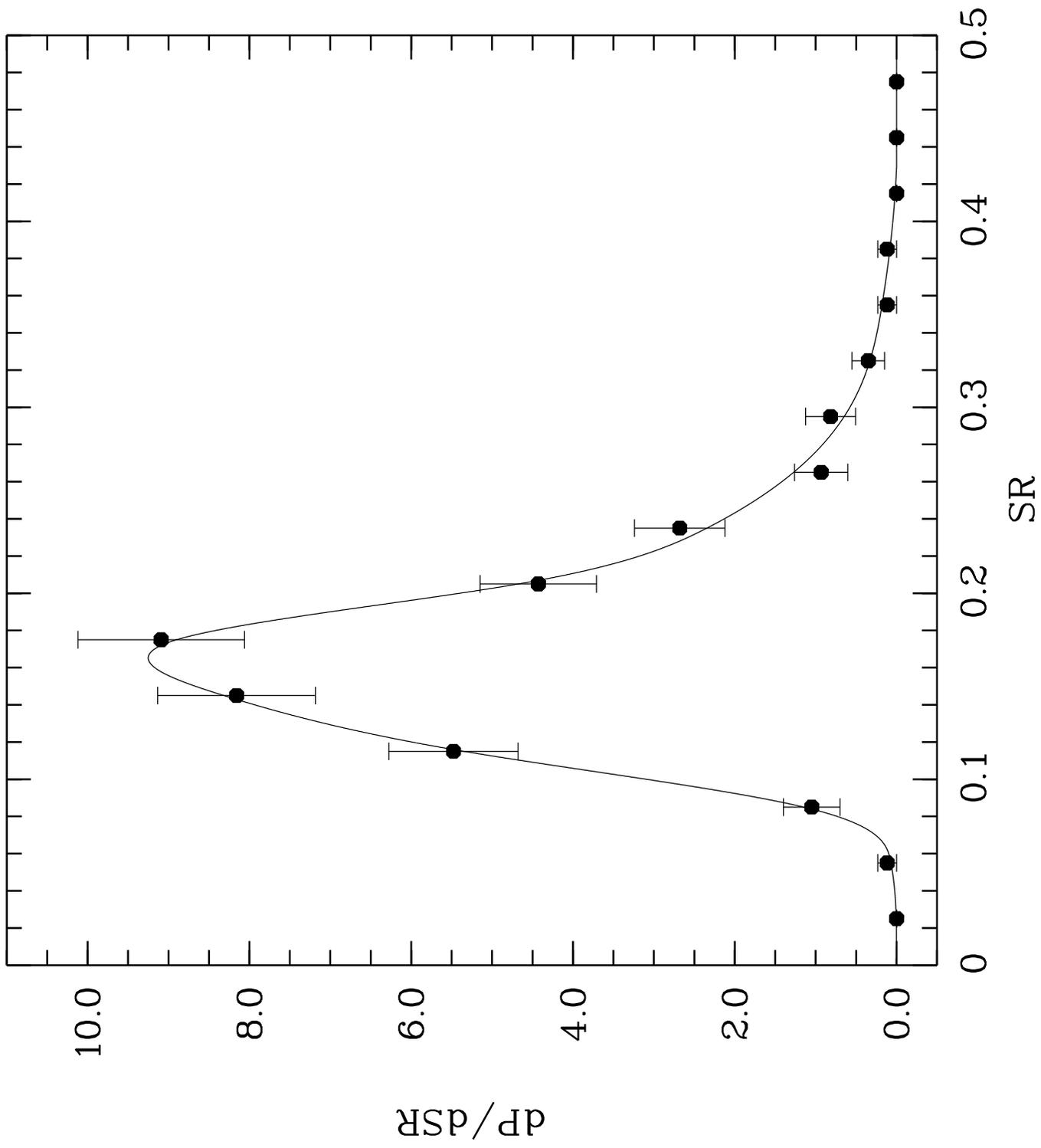
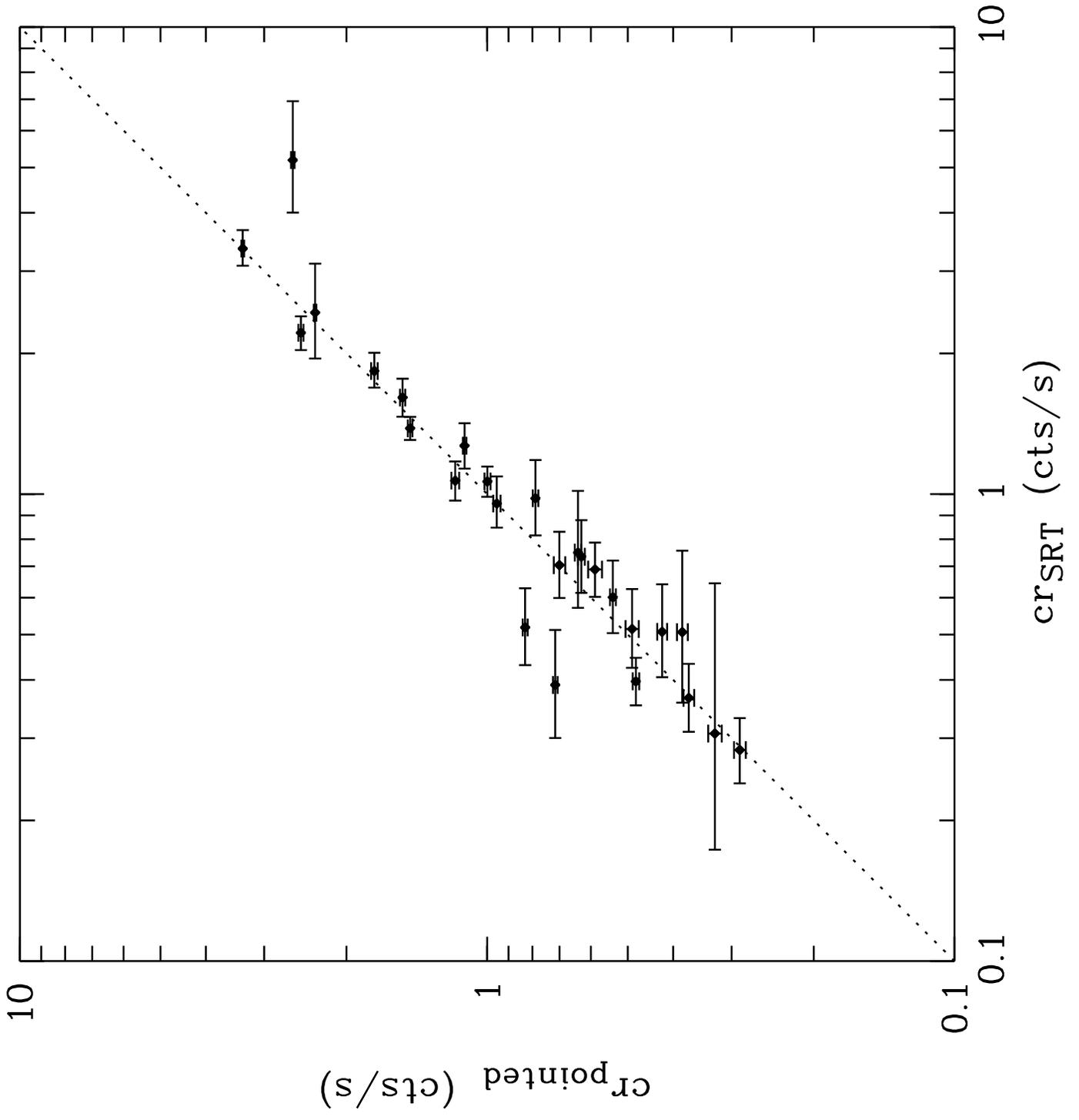
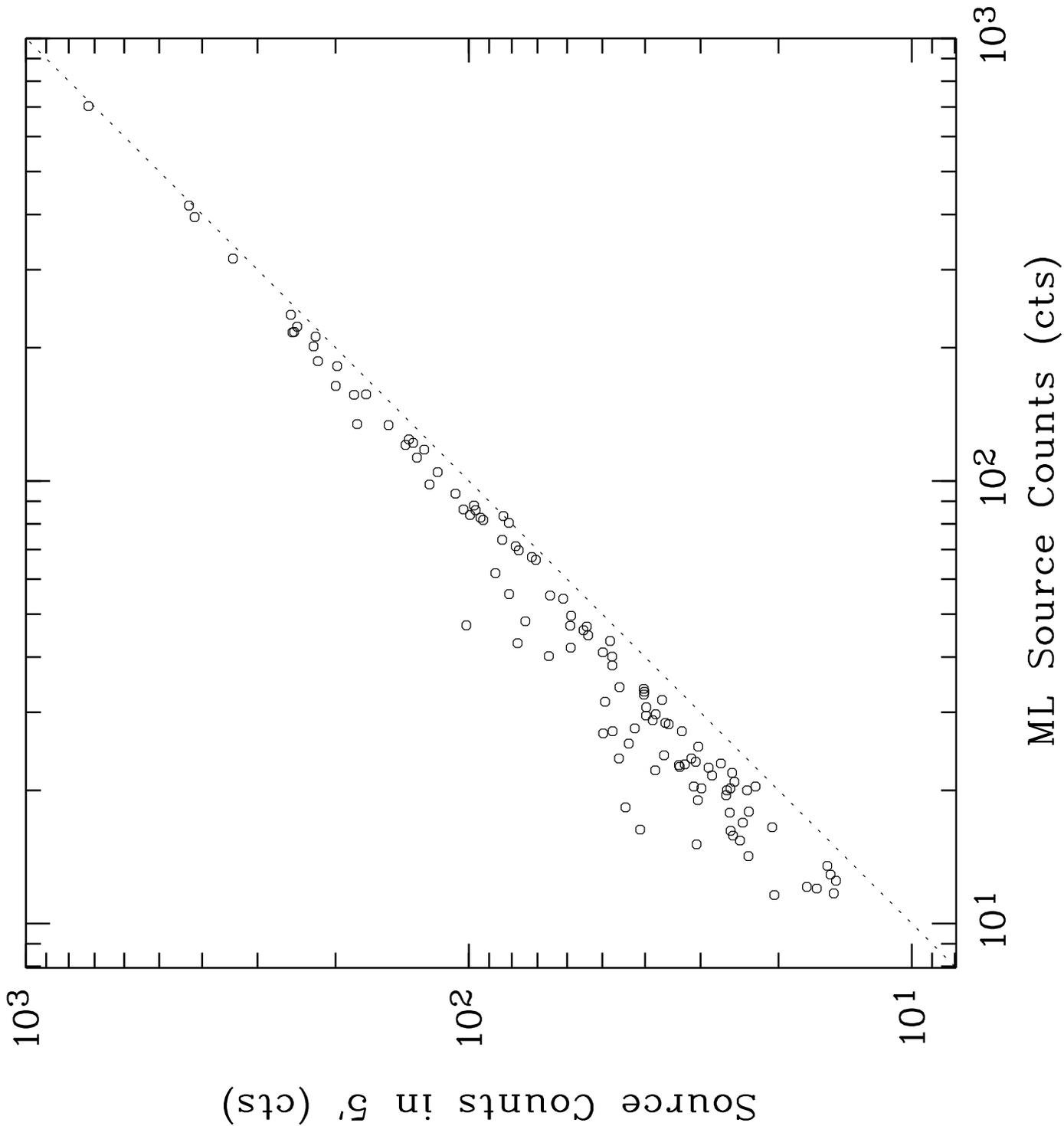
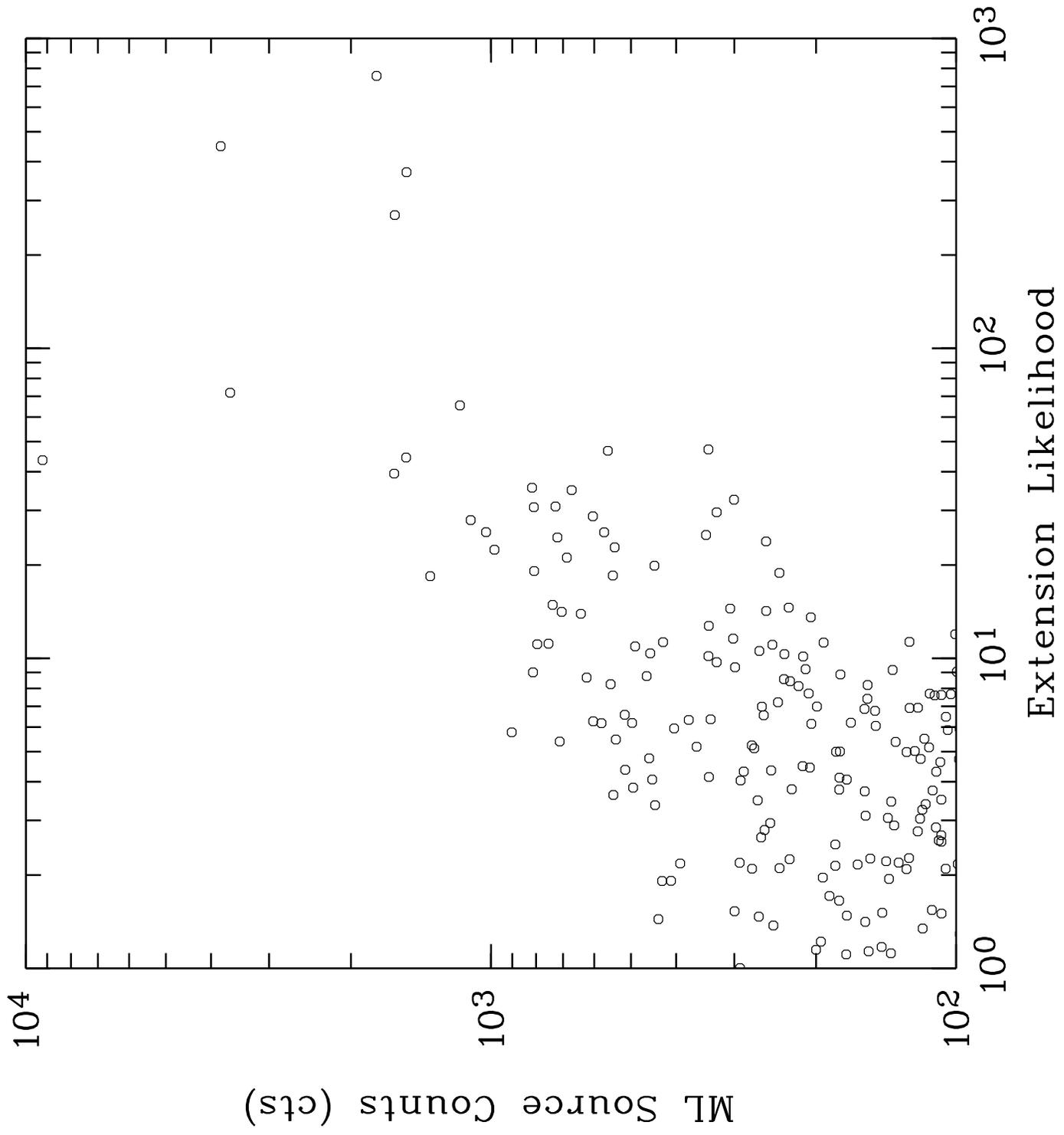
\end{document}